\documentclass[journal]{IEEEtran}
\usepackage{cite}
\usepackage{graphicx,amssymb,lineno}
\usepackage{amsmath,amsfonts,amssymb}

\newtheorem{definition}{Definition}
\usepackage{algorithm}
\usepackage{algorithmic}
\usepackage[usenames]{color}
\usepackage{float}
\usepackage{mathtools}
\usepackage{cite}
\usepackage{booktabs}

\usepackage{graphicx,graphics,color,epsfig,subfigure,graphpap,rotate}
\usepackage{times, verbatim, subfigure, epsfig, graphicx, latexsym, amsmath}
\usepackage{url}
\usepackage{subfigure}

\usepackage{epstopdf}

\makeatletter
\def\ScaleIfNeeded{%
\ifdim\Gin@nat@width>\linewidth \linewidth \else \Gin@nat@width
\fi } \makeatother

\begin{document}

\title{\huge QoS-aware User Association and Transmission Scheduling for Millimeter-Wave Train-ground Communications}
\author{
Xiangfei~Zhang,
Yong~Niu,~\IEEEmembership{Senior Member,~IEEE},
Tao~Yang, Xian~Xiao,
Jianwen~Ding,~\IEEEmembership{Senior Member,~IEEE},
Sheng~Chen,~\IEEEmembership{Fellow,~IEEE},
Zhangdui~Zhong,~\IEEEmembership{Fellow,~IEEE},
Ning~Wang,~\IEEEmembership{Member,~IEEE}
and Bo~Ai,~\IEEEmembership{Fellow,~IEEE}
\thanks{This study was supported by National Key R\&D Program of China (2020YFB1806903); in part by the National Key Research and Development Program under Grant 2021YFB2900301; in part by the National Natural Science Foundation of China Grants 62221001, 62231009, 61725101, 61961130391, and U1834210; in part by the Fundamental Research Funds for the Central Universities (Science and technology leading talent team project) 2022JBXT001; in part by the Fundamental Research Funds for the Central Universities, China, under grant number 2022JBQY004.} %
\thanks{X. Zhang, T. Yang, X. Xiao and J. Ding are with the State Key Laboratory of Rail Traffic Control and Safety, and also with the Frontiers Science Center for Smart High-speed Railway System, Beijing Jiaotong University, Beijing 100044, China (e-mails: zxfei78@163.com, 18211241@bjtu.edu.cn, 18211205@bjtu.edu.cn, jwding@bjtu.edu.cn).} %
\thanks{Y. Niu is with the State Key Laboratory of Rail Traffic Control and Safety, Beijing Jiaotong University, Beijing 100044, China, and also with the National Mobile Communications Research Laboratory, Southeast University, Nanjing 211189, China (e-mail: niuy11@163.com).}
\thanks{Sheng Chen is with the School of Electronics and Computer Science, University of Southampton, Southampton SO17 1BJ, U.K. (e-mail: sqc@ecs.soton.ac.uk).} %
\thanks{Z. Zhong and B. Ai are with the State Key Laboratory of Rail Traffic Control and Safety, and also with the Beijing Engineering Research Center of High-speed Railway Broadband Mobile Communications, Beijing Jiaotong University, Beijing 100044, China (e-mails: boai@bjtu.edu.cn; zhdzhong@bjtu.edu.cn).}
\thanks{N. Wang is with the School of Information Engineering, Zhengzhou University, Zhengzhou 450001, China (e-mail: ienwang@zzu.edu.cn).}
\vspace*{-5mm}
}

\maketitle
\begin{abstract}
With the development of wireless communication, people have put forward higher requirements for train-ground communications in the high-speed railway (HSR) scenarios. With the help of mobile relays (MRs) installed on the roof of the train, the application of Millimeter-Wave (mm-wave) communication which has rich spectrum resources to the train-ground communication system can realize high data rate, so as to meet users' increasing demand for broad-band multimedia access. Also, full-duplex (FD) technology can theoretically double the spectral efficiency. In this paper, we formulate the user association and transmission scheduling problem in the mm-wave train-ground communication system with MR operating in the FD mode as a nonlinear programming problem. In order to maximize the system throughput and the number of users meeting quality of service (QoS) requirements, we propose an algorithm based on coalition game to solve the challenging NP-hard problem, and also prove the convergence and Nash-stable structure of the proposed algorithm. Extensive simulation results demonstrate that the proposed coalition game based algorithm can effectively improve the system throughput and meet the QoS requirements of as many users as possible, so that the communication system has a certain QoS awareness.
\end{abstract}


\section{Introduction}\label{S1}

As a vital part of the transport infrastructure, high-speed railway (HSR) has been developing rapidly in the recent years. By June 2018, the total operational mileages of HSRs in the world has reached 49,000km with additionally more than 16,000km under construction \cite{HSR2018}. Many countries, including China, Japan, Germany and so on, have already operated trains with speeds exceeding 200km/h.

On the other hand, since many passengers are accustomed to broadband wireless access in
their daily
living environment, more and more people hope to
have
high-quality broadband wireless access on mobile terminals.
However, in the HSR scenarios, the high-speed of train causes frequent handovers. When the cell radius is 1 to 2km, the train running at 350km/h will handover every 20 to 40s~\cite{Ai}. In addition, the rapid relative movement between the train and track-side base station (BS) causes more serious Doppler shift and smaller channel coherence time. So the wireless channel in HSR scenarios has obvious non-stationary and fast time-varying characteristics, which seriously reduces the performance of train-ground communication systems~\cite{Doppler}. Moreover, due to the complexity and non-stationarity of HSR scenarios, there are weak field strength and blind areas, and the train body of metal material causes great penetration loss to the signal from the BS~\cite{HSRMR}.
Therefore,
meeting
passengers'
compelling
demand for broadband mobile communications
in the HSR environment has
become a key technical
challenge.
It
has become
particularly important to carry out research on broadband wireless communication technologies
in HSR scenarios. However, the current widely-used communication technologies under HSR scenarios, such as Global System for Mobile Communications for Railways (GSM-R), can
only
support a rate that is too low to meet user's increasing demands for multimedia applications, and the current demand rate of each train is about 37.5Mbps~\cite{xiangfei21}. With the growth of
business
and entertainment
activities
and quality of service (QoS) demands,
the demand rate
may
easily
reach 0.5--5Gbps in the
near
future~\cite{ROF}. Obviously, the current wireless transmission scheme
will be inadequate to
satisfy
the needs of HSR passengers.

To this end, industry and academia have started to exploit the 30-300 GHz millimeter-wave (mm-wave) frequency band. Owing to its wide bandwidth and rich spectrum resources, it is not only consistent with the current development trend of wireless communication systems, but also able to meet the demand of multi-gigabit wireless services such as online gaming, video calling and so on~\cite{J1}, which the current communication systems for railways like GSM-R and Long Term Evolution for Railways (LTE-R) can't provide. Thus, mm-wave communications are regarded as a promising technology to meet the growing network capacity requirements caused by the rapid development in the HSR industry~\cite{channel}\cite{J8}.

But compared with the communication systems of lower frequency band, mm-wave communication suffers from greater propagation loss. Because the free space path loss (FSPL) is proportional to the square of the carrier frequency, the FSPL in 60 GHz is 28 decibels (dB) higher than that in 2.4GHz~\cite{24ghz}. In order to compensate for strong link attenuation, mm-wave communication uses directional antenna to achieve beamforming for improving antenna gain. At the same time, due to the short wavelength of mm-wave and the relatively small size of the antenna array, we can deploy mobile relays (MRs) on the roof of the train, which can overcome the huge penetration loss caused by the train body. Furthermore, the full-duplex (FD) communication technology allows wireless communication devices to simultaneously transmit and receive signals on the same frequency band, and the capacity of the communication system can be doubled theoretically~\cite{FDintr}. Therefore, FD technology is regarded as one of the key physical layer technologies of 5G. The biggest challenge of FD communication is the elimination of self interference (SI), which is caused by the signal leakage from the sending end to the corresponding receiving end~\cite{fd1}~\cite{fd2}.

Therefore, it would be desirable for the MRs to adopt both the mm-wave and FD technology, and simultaneously serve HSR passengers and ground users. The track-side BS allocates part of the bandwidth to the MRs, allowing some users to directly communicate with the track-side BS in a traditional manner, and other users to be associated with the MRs to obtain wireless access. At this time, determining the user association mechanism in the investigated train-ground communication system is a key challenge, and proposing an effective user association scheme to maximize the number of users whose QoS requirements are satisfied is the focus of this paper. If mm-wave communication and FD technology can be applied to the train-ground communication system in the HSR scenarios, high-rate transmission can be achieved between the BS and MR to significantly improve the communication service performance~\cite{LuExploring2018}.

However, the current research on the application of mm-wave communication and FD technology to train-ground communications are very limited. In this paper, we study the problem of user association in the mm-wave train-ground communication system, in which the MR adopts FD mode. We focus on the cooperation between track-side BS and FD MR in user association to optimize system performance, especially to maximize the system throughput and the number of users meeting QoS requirements. To this end, we propose a user association optimization scheme based on coalition game theory. In this scheme, each player will try its best to achieve the maximum system throughput. Then we propose a transmission scheduling algorithm to maximize the number of users whose QoS requirements are satisfied. The main contributions made in this paper are summarized as follows.
\begin{itemize}
\item Mm-wave communications and FD technology are combined to improve the capacity of  train-ground communication system in HSR scenarios.
\item A low complexity user association and transmission scheduling algorithm is proposed to reduce the implementation complexity while ensuring the network capacity is at a high level.
\item We focuse on users' QoS requirements to improve user communication experience in HSR scenarios.
\end{itemize}

Firstly, HSR communications are still based on GSM-R at present. The mm-wave band with rich spectrum resources can fundamentally solve the problem of lack of wireless resources in HSR scenarios, and FD technology has the potential to double communication capacity in theoretically. Therefore, we combined mm-wave communications and FD technology to build a new train-ground wireless communication framework to significantly improve transmission performance in HSR scenarios.

Secondly, considering the fast time-varying of channels and frequent handoff in HSR scenarios, the optimization algorithm that require complex iteration is no longer applicable to the communication problem in HSR scenarios. Therefore, we proposed a low complexity user association algorithm based on coalition game theory, and theoretically prove its convergence and stability. At the same time, we proposed a greedy algorithm for transmission scheduling to further improve capacity performance. The proposed user association and transmission scheduling algorithms both ensure the system performance and reduce the implementation complexity, making them more suitable for application in HSR scenarios.

Thirdly, different from the most existing user association researches, we pay more attention to the users' QoS requirements in the HSR scenarios. Our goal is to schedule as many data flows as possible while meeting the users' QoS requirements, rather than just to increase system capacity, and the proposed algorithm will effectively improve the user communication experience in the HSR scenarios.

The rest of the paper is organized as follows. In Section~\ref{S2}, we provide an detailed overview of the related work. Section~\ref{S3} introduce the train-ground communication system model, and Section~\ref{S4} formulate the problem of user association. The proposed user association scheme based on coalition game theory and transmission scheduling algorithm are presented in Section~\ref{S5} and evaluated in Section~\ref{S6}. The conclusions of this paper are drawn in Section~\ref{S7}.

\section{Related Work}\label{S2}

We
can classify
the related works into three categories: (a) mm-wave communications in wireless networks; (b) FD communications in wireless networks; and (c) application of game theory into resource allocation. We examine these related work in this section.

\paragraph{Mm-wave Communications}
There has been considerable research on mm-wave based wireless communications and networks. Most of these researches focus on 28GHz, 38GHz, 60GHz and E-band (71-76GHz and 81-86GHz). The rapid development of complementary metal oxide semiconductor (CMOS) in radio frequency integrated circuits has paved the way for electronic products in mm-wave band~\cite{DoanDesign2004}. At the same time, several international standards for indoor wireless personal area network (WPAN) and wireless local area network (WLAN), such as ECMA-387, IEEE 802.15.3c and IEEE 802.11ad, have been developed.

In terms of the propagation characteristics of mm-wave, Singh \emph{et al.} mentioned that the FSPL of 60GHz mm-wave is 28dB higher than that of 2.4GHz~\cite{kt15}. At the same time, the authors of~\cite{kt14} pointed out that mm-wave is more sensitive to being blocked by obstacles. For these problems, considering that high-speed trains (HSTs) generally operate in rural areas and viaducts, the buildings in the HSR scenarios are short and sporadic, which reduces the probability of mm-wave being blocked. In addition, when the MR is installed on the roof of the train, the directional characteristics of mm-wave can be further utilized, while avoiding the huge penetration loss caused by the train body~\cite{kt18}~\cite{kt19}. Theiss \emph{et al.} analyzed the propagation characteristics of the mm-wave in the HSR scenarios, and showed the measurement results from the mm-wave transceiver prototype in the Japanese Shinkansen environment~\cite{kt20}. Subsequently, He \emph{et al.} analyzed the influence of many typical objects in the HSR scenarios on mm-wave propagation, obtained the corresponding electromagnetic parameters, and established a mm-wave channel model closed to the real train-ground communications~\cite{kt23}. Yuan \emph{et al.} proposed a profit-maximized collaborative computing offloading and resource allocation algorithm based on an architecture including terminal, edge computing, and cloud data centers (CDC) layers, but the introduction of edge computing into the mm-wave train ground communication system may increase the end-to-end delay~\cite{refr21}. However, due to the difficult propagation characteristics of mm-wave and the particularity of HSR scenarios, there are still many challenges ahead.

\paragraph{FD Communications}
FD wireless communications have also attracted much efforts in the research community in the past decade. For example, Chen \emph{et al.} uses the FD jamming receiver to assist the multi-antenna transmitter to realize covert communications, and the proposed scheme maximizes the system throughput while ensuring covertness~\cite{FDchen}. Cao \emph{et al.} explored the downlink non-orthogonal multiple access system assisted by multi-antenna FD relay, and proposed a two-stage interference scheme, which significantly improved the security performance~\cite{FDcao}. Li \emph{et al.} combined FD technology and physical layer coding in the two-way relay channel, and proposed two transmission schemes, which improved the system throughput and reduced the time slot required for unit information transmission~\cite{FDli}. Cheng \emph{et al.} proposed a request to send and FD clear to send based media access control (MAC) protocol, which can support two-way and one-way links in a wireless FD network, achieving higher normalized system throughput~\cite{FDcheng}. In addition, Kai \emph{et al.} formulated the joint optimization problem of tarnsmission scheduling and power control in a three-node FD WLAN, and proposed an efficient algorithm based on alternating optimization and successive convex approximation with low complexity~\cite{FDkai}.

The toughest challenge of FD communications is the cancellation of SI caused by the signal leakage from the transmitter of a terminal to the receiver of the same terminal~\cite{WenTime2016}. In order to realize the high spectral efficiency, the SI has to be reduced by 110dB with the help of antenna separation, analog-domain and digital-domain cancellation~\cite{DingQoS2018}. At the same time, due to the directional transmission characteristics of mm-wave communications, the directivity of the transmitting and receiving beams can naturally reduce the SI \cite{NiuA2015}. In order to achieve the theoretical double system throughput, Nidal \emph{et al.} proposed a scheme to restrain SI by using a reflective array~\cite{FDnidal}.

The related work show that FD communications have great research value in cellular networks, WiFi networks, mm-wave networks and heterogeneous networks. However, there are still few applications of FD in HSR scenarios. In the train-ground communication system model established in this paper, we exploit the advantages of FD technology and mm-wave communications, which can jointly bring about huge improvements on network capacity and user experience.


\paragraph{Game Theory based Resource Allocation}
Considering the inherent problems such as fast time-varying of channels and frequent handoff in HSR scenarios, the optimization algorithms that used to require complex iteration are no longer applicable to the train-ground communications in HSR scenarios, especially once new users enter the system, the previous experience will lose its effectiveness. To address the dual requirements of normal and quick-response ISL schedulings, a data-driven heuristic assisted memetic algorithm (DHMA) is proposed by Du \emph{et al.}. The main idea of it is to address normal and quick-response schedulings separately, while high-quality normal scheduling data are also trained for quick-response use~\cite{refr22}. However, this method relies on large data sets, and mm-wave communication has not been applied to the train-ground communication system in HSR scenarios at present, so the lack of data sets makes this method not applicable in HSR scenarios. To improve the real-time performance of networked control systems, the joint problem of message fragmentation and no-wait scheduling is explored by Jin \emph{et al.}~\cite{refr23}. The proposed algorithm is based on optimization module theory, which is not suitable for train-ground communication in HSR scenarios with fast time-varying channels because of its high complexity. To provision high-speed Internet services in HSTs, Taheri \emph{et al.} proposed a novel solution named Free-space OptiCs Utilization in high-Speed trains (FOCUS). However, the authors did not consider meeting the QoS requirments of different users~\cite{refr24}.

Game theory has been widely applied in wireless networks to improve the utilization of limited resources. For example, the authors of \cite{LiConcurrent2017} proposed a time resource sharing scheme based on Stakelberg game, in which the D2D links that cause interference can only access the time resources by paying a higher price, and it can realize distributed power control to reduce interference and improve the network throughput. Paik \emph{et al.} formulated the downlink resource allocation problem in the multi-cell OFDMA system as a non-cooperative game, and proposed a coalition game based scheme using adaptive modulation and coding to maximize system capacity~\cite{GAMEpaik}. Wang \emph{et al.} proposed a novel resource allocation algorithm based on game theory in the full-frequency multiplexing OFDMA system, and established a non-cooperative game model based on the arctangent function, which achieved lower complexity while ensuring network capacity~\cite{GAMEwang}. The authors of~\cite{GAMErathi} discussed the influence of Stackelberg game on system performance when optimizing radio resource allocation in heterogeneous networks, and the established game theory model provides guidance for exploring more efficient resource allocation schemes.




\begin{figure}[t]
\begin{center}
\includegraphics[width=0.95\columnwidth,height=1.7in]{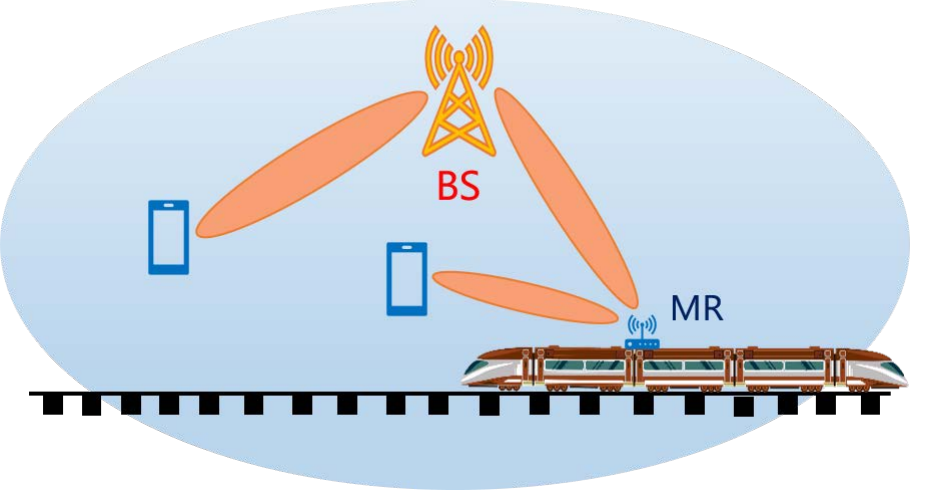}
\end{center}
\caption{Illustration of an mm-wave train-ground communication system with MR operating in FD mode.}
\label{System} 
\end{figure}

It can be concluded that in the existing works, there is a paucity of contributions on user association for mm-wave train-ground communication systems with FD MR as well as transmission scheduling of data flows in HSR scenarios. These motivate our research.

\section{System Model}\label{S3}

%

This paper considers a mm-wave train-ground communication system using FD MR, as shown in Fig.~\ref{System}. In this model, the track-side BS reserves part of the bandwidth resources, and the remaining bandwidth is allocated to MR, and the two parts of bandwidth resources are mutually exclusive. All the devices in this train-ground communication system work in the mm-wave frequency band. In addition, both the BS and MR are equipped with directional antennas, and transmit and receive beamforming are allowed to achieve higher mm-wave directional gain. Ground users and passengers in the HST can choose to directly communicate with the BS in a traditional manner, or use MR to achieve two-hop transmission, so there are problems with user association and transmission scheduling. Assuming that the bandwidth allocation ratio of track-side BS and MR has been predetermined, we hope to find the optimal user association mechanism to achieve high network capacity and make the number of users meeting the QoS requirements as many as possible.


\begin{figure}[t]
\begin{center}
\includegraphics[width=0.85\columnwidth,height=1.1in]{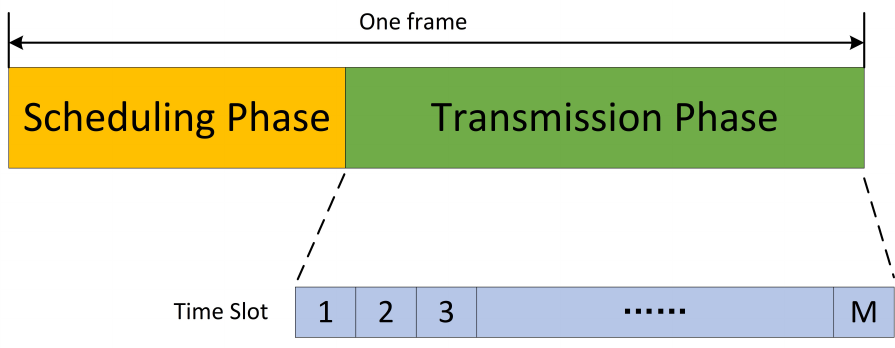}
\end{center}
\caption{The structure of one superframe.} \label{Superframe}
\end{figure}

We adopt the frame structure of the MAC layer \cite{SonOn2012} which is illustrated in Fig.~\ref{Superframe}. Each superframe consists of a scheduling phase and a transmission phase. During the scheduling phase, the BS or MR receives the QoS request for each flow and makes the scheduling decision, which is then broadcasted to the entire network. During the transmission phase, the frame is further divided into $M$ equal time slots. At each time slot, the data flow can be transmitted based on scheduling decisions made during the scheduling phase. Depending on the user association and scheduling decision, some users in the system receive services through the BS, and others receive services through the MR. Multiple users associated with BS or MR use the TDMA mode to access the network. Considering that each user has certain QoS request, which is described by its throughput requirement in this paper, the BS and MR work jointly to satisfy the user's QoS request through transmission scheduling.

First of all, we assume that there are $N$ flows requesting transmission time slots in the superframe. Since each time slot can only be occupied by one flow, we have $N\le M$. For flow $i$, we denote its sender and receiver by $s_i$ and $r_i$, respectively. The distance between $s_i$ and $r_i$ is denoted by $l_i$. We also denote the transmitting antenna gain of $s_i$ in the direction of $s_i$ to $r_i$ by $G_t(s_i,r_i)$, and the receiving antenna gain of $r_i$ in the direction of $r_i$ to $s_i$ by $G_r(r_i,s_i)$. According to the path loss model \cite{CaiRex2010}, the received signal power at $r_i$ for the signal transmitted from $s_i$ to $r_i$ can be calculated as
\begin{align}\label{P_t} 
P_r(i) =  k_0 \cdot G_t(s_i,r_i) \cdot G_r(r_i,s_i) \cdot l_i^{-n} \cdot P_t(i) ,
\end{align}
where $P_t(i)$ is the transmission power of flow $i$, $k_0$ is a constant that is proportional to $\left(\lambda /4\pi\right)^2$ with $\lambda$ denoting the signal wavelength, and $n$ is the path loss exponent~\cite{CaiRex2010}~\cite{QiaoSTDMA2012}. Then the received signal-to-noise power ratio (SNR) of flow $i$ can be calculated as
\begin{align}\label{SNR_i} 
\text{SNR}_i = \frac{P_r(i)}{N_0 W} = \frac{k_0 \cdot G_t(s_i,r_i) \cdot G_r(r_i,s_i) \cdot l_i^{-n} \cdot P_t(i)}{N_0 W} ,
\end{align}
where $W$ is the bandwidth, and $N_0$ is the unilateral noise power spectral density of the Gaussian channel~\cite{QiaoSTDMA2012}. Therefore the achievable data rate of flow $i$ according to the Shannon's channel capacity can be expressed as
\begin{align}\label{R_i} 
R_i = \eta \cdot W \cdot \log_2\left(1+\frac{P_r(i)}{N_0 W}\right) ,
\end{align}
where $\eta\in (0, ~ 1)$ describes the efficiency of the transceiver design. If the number of time slots allocated to flow $i$ is $\delta_i$, the achieved throughput of flow $i$ can be expressed as
\begin{align}\label{q_i} 
q_i = \frac{R_i \cdot \delta _i}{M} .
\end{align}
where $M$ denotes the number of time slots of the transmission phase.

\begin{figure}[t]
	\begin{center}
		 \includegraphics*[width=0.6\columnwidth,height=1.33in]{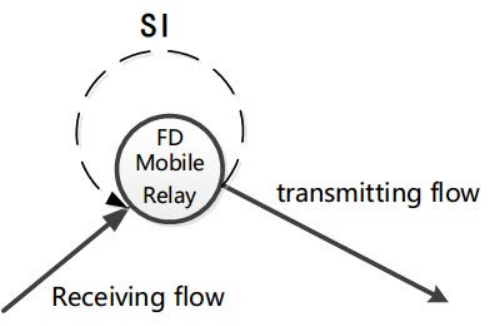}
	\end{center}
	\caption{SI at an FD MR.} \label{SI}
\end{figure}


Because the MR adopts the FD mode, it causes a certain amount of SI (i.e., due
to imperfect SI cancellation). As shown in Fig.~\ref{SI}, the so-called SI refers to the interference at the MR receiver caused by its own transmitter. Let the MR's transmission power be $P_t$, then the SI power can be modeled as $\beta \cdot P_t$, where $\beta$ represents the SI cancellation level. Therefore, if receiver $r_i$ of flow $i$ is the MR with FD mode, then the achievable data rate of flow $i$ should be modified as
\begin{align}\label{M-R_i} 
R_i = \eta \cdot W \cdot \log_2\left(1+\frac{P_r(i)}{N_0 W + \beta \cdot P_t(i)}\right) .
\end{align}


\section{Problem Formulation}\label{S4}

In this paper, efficient transmission scheduling is achieved by maximizing the number of users whose QoS requests are met. Therefore, we establish an optimization problem with user association and user time slot scheduling as decision variables, aiming to maximize the number of users who meet the condition. As aforementioned, there are $N$ flows to be transmitted with $M$ available time slots. The scheduler partitions the superframe into $K$ pairings, where a pairing is defined as several consecutive time slots. For each flow $i$, we define a binary variable $\alpha^k_i$ to indicate whether flow $i$ is scheduled in the $k$th pairing. Specifically, if flow $i$ is scheduled in the $k$th pairing, $\alpha^k_i=1$, otherwise $\alpha^k_i=0$. The number of time slots in the $k$th pairing is denoted by $\gamma^k$.

\begin{figure}[t]
	\begin{center}
		 \includegraphics*[width=0.85\columnwidth,height=2.1in]{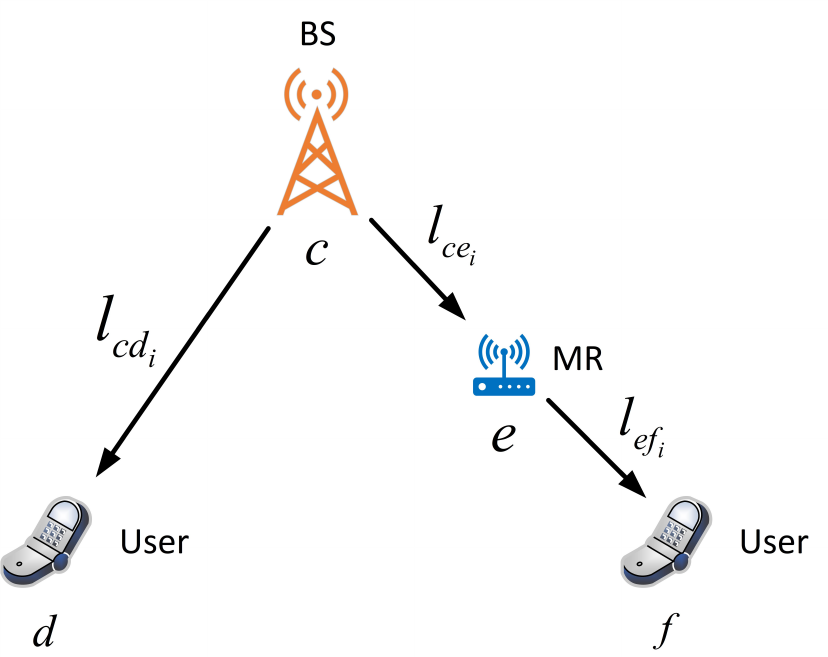}
	\end{center}
	\caption{Schematic diagram of link model.} \label{model}
\end{figure}

Given the total bandwidth $W$, the BS is allocated with the bandwidth $a \cdot W$ and the MR with the bandwidth $b \cdot W$, where $a+b=1$. Further define two binary variables $\mu_i$ and $\lambda_i$, and $\mu_i$ indicates whether flow $i$ is associated with the BS. specifically, $\mu_i=1$ indicating that flow $i$ is associated with the BS, otherwise $\mu_i=0$. Similarly, if flow $i$ is associated with the MR, the indicator $\lambda_i=1$, otherwise $\lambda_i=0$.
To show how we calculate the transmission rates of different links, consider Fig.~\ref{model} as an example.


In the example shown in Fig.~\ref{model}, the transmission rate over the link $l_{cd_i}$ that flow $i$ is sent directly from BS $c$ to user $d$ can be expressed as
\begin{align}\label{R^k_{iBS}} 
\begin{array}{l}
R_{BS,i}^k = R_{i(c \to d)}^k\\
 ~~~~~~~= \eta  \cdot a \cdot W \cdot {\log _2}\left( {1 + \frac{{\alpha _i^k{k_0}{G_t}(c,d){G_r}(d,c)l_{c{d_i}}^{ - n}{P_t}(i)}}{{{N_0} \cdot a \cdot W}}} \right) .
\end{array}
\end{align}


Taking into account the RSI at FD MR, the transmission rate over the link $l_{ce_i}$ that flow $i$ is sent from BS $c$ to MR $e$ can be expressed as
\begin{small}
\begin{align}\label{R^k_{i1}} 
R^k_{i(c\rightarrow e)} = \eta \cdot b \cdot W\log_2\! \left(\! 1\! +\! \frac{\alpha^k_i k_0 G_t(c,e) G_r(e,c) l_{ce_i}^{-n} P_t(i)}{N_0 \cdot b \cdot W + \beta \cdot P_t(i)}\!\right)\! .
\end{align}
\end{small}

Similarly, it is assumed that the transmission power of BS and MR is $P_t$, then the transmission rate over the link $l_{ef_i}$ that flow $i$ is sent via MR $e$ to user $f$ can be expressed as
\begin{small}
\begin{align}\label{R^k_{i2}} 
R^k_{i(e\rightarrow f)}\! = \eta \cdot b \cdot W \cdot \log_2\! \left(\! 1\! +\! \frac{\alpha^k_i k_0 G_t(e,f) G_r(f,e) l_{ef_i}^{-n} P_t(i)}{N_0 \cdot b \cdot W + \beta \cdot P_t(i)}\! \right)\! .
\end{align}
\end{small}
Hence, the rate of flow $i$, sent from BS via MR to user, is
\begin{align}\label{R^k_{iMR}} 
R^k_{{\rm MR},i} = \min\big\{R^k_{i(c\rightarrow e)}, R^k_{i(e\rightarrow f)}\big\}.
\end{align}

Let the required throughput for flow $i$ be $Q_i$, and denote the actual throughput for flow $i$ as $q_i$, therefore we have
\begin{align}\label{A-q_i} 
q_i = \frac{\sum_{k=1}^K\left(\mu_i \cdot R^k_{{\rm BS},i} + \lambda_i \cdot R^k_{{\rm MR},i}\right) \cdot \gamma_k}{M} .
\end{align}

We further define a binary variable $I_i$ to indicate whether the actual throughput $q_i$ achieved for flow $i$ meets the QoS requirement $Q_i$, specifically,
\begin{align}\label{A-I_i} 
I_i =& \left\{\begin{array}{cl}
1 , & q_i \ge Q_i , \\
0 , & q_i < Q_i .
\end{array} \right.
\end{align}
where the QoS requirement here is defined as the minimum throughput requirement of the data flow.

Given the QoS requirements $Q_i$ for all the $N$ flows, $1\le i\le N$. Since the number of time slots $M$ in the transmission phase is limited, the optimal scheduling should accommodate as many flows as possible. Therefore, our goal is to maximize the number of flows that satisfy their QoS requirements, and the objective function can be expressed as
\begin{align}\label{Opt-I} 
\max\sum_{i=1}^N I_i .
\end{align}

Next we analyze the constraints of this optimal transmission scheduling. First, each pairing of the superframe can only be occupied by a flow once, and therefore
\begin{align}\label{Cont-1} 
\sum_{k=1}^K{\alpha}^{k}_{i}=1, ~ \forall i .
\end{align}
Second, there are only $M$ time slots in the superframe, so we have
\begin{align}\label{Cont-2} 
\sum_{k=1}^K \gamma^k \leq M .
\end{align}
Third, flow $i$ can only be associated with either the BS or the MR, not the both, that is,
\begin{align}\label{Cont-3} 
\left\{ {\begin{array}{*{20}{c}}
{{\mu _i} + {\lambda _i} = 1}\\
{{\mu _i}\cdot{\lambda _i} = 0}
\end{array}}\right.{\rm{,}}~~\forall i.
\end{align}

\begin{figure*}[t]
\vspace*{-2mm}
\begin{center}
\includegraphics[width=0.75\linewidth]{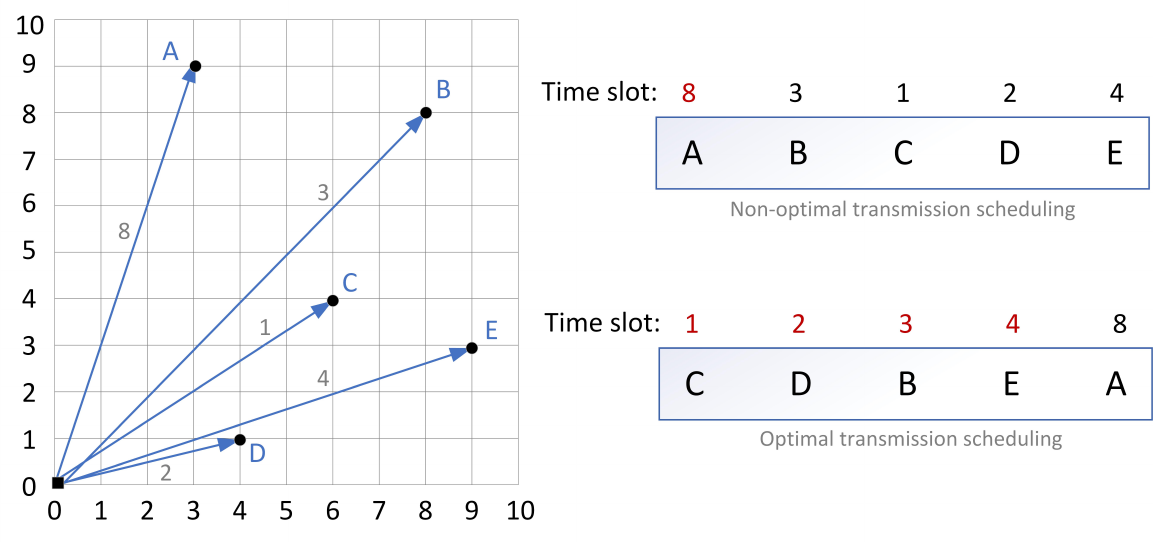}
\end{center}
\vspace*{-4mm}
\caption{Comparison of non-optimal and optimal transmission scheduling. The system has $M=10$ time slots to schedule $N=5$ flows.}
\label{example} 
\vspace*{-3mm}
\end{figure*}

We utilize the simple example as shown in Fig.~\ref{example} to illustrate the basic idea of our optimal transmission scheduling strategy. There are five nodes uniformly distributed in a square area of $10\,\text{m}\times 10\,\text{m}$, and they are denoted by $A$, $B$, $C$, $D$ and $E$, respectively. There are $M=10$ time slots in the superframe, and $N=5$ flows in the network to be scheduled. To meet the throughput requirements, 8 time slots are needed for the flow $O\to A$, 3 time slots for the flow $O\to B$, 1 time slot for the flow $O\to C$, 2 time slots for the flow $O\to D$, and 4 time slots for the flow $O\to E$. Without transmission scheduling and simply carrying out the transmission scheduling for the 5 flows in the order of $A$, $B$, $C$, $D$, and $E$, only one flow, $A$, can meets its QoS requirement. To maximizing the number of flows that can meet their QoS requirements, the flow scheduling should be carry out according to the required time slots for meeting the required throughput in the order of low to high, that is, the transmission scheduling should be $C$, $D$, $B$, $E$ and $A$. It can be seen that with this transmission scheduling, four flows, namely $B$, $C$, $D$ and $E$, can meet their QoS requirements.

Considering the height of the MRs, the link between the MRs and the ground BS can be regarded as line-of-sight (LOS) transmission, and the railway is generally located in spacious scenarios, so it is considered that the blockages are mainly caused by human body. We consider that in the investigated mm-wave train-ground communication system, the blockage problem caused by human body can be solved in three steps. In the first step, we connect the users to the nearest BS or MR, and propose an average outage probability ${P_b}$ for all links, which note that each link has a random probability of experiencing blockage. In the second step, if the communication link between the user and the associated equipment experiences blockage, this user will re-establish an association relationship with the second closest equipment to actively avoid blockage. In the third step, we study the scenario of non-LOS (NLOS) transmission, which can be enhanced with the assist of intelligent reflecting surface (IRS) and Device-to-Device (D2D) communications. In this paper, for analytical simplicity, it is assumed that the link blockage probability remains stable in a road segment. The link blockage probability ${P_b}$ on average is deemed as a constant in a certain section along the rail track~\cite{blockconstant}.


In summary,
the problem of optimal user association and transmission scheduling, denoted by P1, can be formulated as follows.
\begin{align}
\mbox{(P1)} \;\;
\max &~~(1-P_b)\sum\limits_{i = 1}^N {{I_i}} \label{P1}
\\
\mbox{s.t.} &~~\mbox{Constraints~(\ref{Cont-1}) to~(\ref{Cont-3})}. \nonumber
\end{align}

Because
the data transmission rate of each flow includes a logarithmic function,
the objective function is nonlinear.
So P1 is a mixed integer nonlinear programming problem with binary variable $I_i$. Considering that the user's location, QoS requirements, and the gain of the channel formed by connecting with BS and MR will jointly decide the best user association scheme in the established train-ground communication system, it is more complex than the 0-1 knapsack problem~\cite{01pack}, so P1 is an NP-hard problem, and difficult to solve in polynomial time~\cite{DingQoS2018}.
In the next section,
we introduce
a coalition game theory based algorithm to
solve problem (P1) with competitive solutions.

\section{Coalition Game based Algorithms}\label{S5}

In this section, we will introduce the coalition game from the view point of game theory to solve the problem of user association. Coalition formation game is a subcategory of the cooperative game theory, which is able to determine how users cooperate to achieve the optimal system throughput. After that, a sorting algorithm is proposed to solve the problem of optimal transmission scheduling.

\subsection{Coalition Formation: Model}\label{S5.1}

To tackle the optimization problem formulated in the Section~\ref{S4}, we will introduce a coalition game model, where the flows form coalitions to improve the system utility. In our investigated system, there are $N$ flows, and the flows can choose to associate either with the BS or with the MR. Hence, there are two coalitions, and we denote them by $\mathcal{F}_1$ and  $\mathcal{F}_2$. Specifically, $\mathcal{F}_1$ represents the set of flows associated with the BS, while $\mathcal{F}_2$ represents the set of flows associated with the MR. Clearly, for our entire coalition  $\mathcal{F}=\{\mathcal{F}_1,\mathcal{F}_2\}$, we have
\begin{equation}
{{\cal F}_1} \cap {{\cal F}_2} = \emptyset ,
\end{equation}
\begin{equation}
\left| {\cal F} \right| = \left| {{{\cal F}_1}} \right| + \left| {{{\cal F}_2}} \right| = N ,
\end{equation}
where $\big|\mathcal{F}\big|$ denotes the cardinality of the set $\mathcal{F}$.

Recall that the transmission rate $R^k_{{\rm BS},i}$ over the link that flow $i$ is sent from the BS to a user is given by (\ref{R^k_{iBS}}).
Thus the sum rate associated with the coalition $\mathcal{F}_1$ is
\begin{align}\label{R(F_1)} 
R\big(\mathcal{F}_1\big) = \sum_{i=1}^N R^k_{{\rm BS},i}.
\end{align}

Also the transmission rate $R^k_{{\rm MR},i}$ over the link that flow $i$ is sent via the MR to a user is given by (\ref{R^k_{i1}}) to (\ref{R^k_{iMR}}).
Hence the total throughput associated with the coalition $\mathcal{F}_2$ is
\begin{align}\label{R(F_2)} 
R\big(\mathcal{F}_2\big) =& \sum_{i=1}^N R^k_{{\rm MR},i}.
\end{align}

$R\big(\mathcal{F}_1\big)$ and $R\big(\mathcal{F}_2\big)$ represents the sum profits contributed by the entire coalition $\mathcal{F}$ with the particular coalition partition of $\mathcal{F}_1$ and $\mathcal{F}_2$. Hence, the user association model with resource sharing relations for our investigated mm-wave train-ground communication system is established in a coalition game with transferable utility. In this model, the data flows, as players, tend to form a coalition to associate either with the BS or with the MR, in order to maximize the total profit of the system.

\subsection{Coalition Formation: Algorithm}\label{S5.2}

In this section, We will propose a coalition formation algorithm based on the coalition formation game. One key point to coalition formation is what strategy to adopt by each flow. An efficient coalition formation algorithm should be able to compare the sum profits of the coalitions and to find a suitable coalition for each flow to join so as to achieve the optimization goal. Thus, we first introduce the concept of preference order to enable the comparison of the profits for different coalitions.

\begin{definition}[{preference order}]\label{D1}
For any flow $i\in \mathcal{N}=\{1,2,\cdots , N\}$, the preference order $\succ_i$ is defined as a complete, reflexive, and transitive binary relation over the set of all the coalitions that flow $i$ can possibly form.
\end{definition}

Therefore, the flows in our coalition game have the right to choose to join or leave a coalition according to their preference order, and they tends to join a coalition based on which it prefers to being a member. For any given flow $i\in \mathcal{N}$, $\mathcal{F}_c\succ_i \mathcal{F}_{\bar{c}}$ implies that flow $i$ is more willing to be a member of the coalition $\mathcal{F}_c$, that is, $i\in \mathcal{F}_c$, than with $\mathcal{F}_{\bar{c}}$, which does not include the case that flow $i$ prefers these two coalitions equally. In different applications, the preferences for flows can be quantified in different way. In this paper, we adopt the utilitarian preference order \cite{SaadCoalitional2009}.

\begin{definition}[{utilitarian preference order}]\label{D2}
For any flow $i\in \mathcal{N}=\{1,2,\cdots , N\}$, the utilitarian preference order is defined by the following equivalence:
\begin{small}
\begin{align}\label{F_c} 
\mathcal{F}_c\succ_i \mathcal{F}_{\bar{c}}\Leftrightarrow R\big(\mathcal{F}_c\big) + R\big(\mathcal{F}_{\bar{c}}\setminus i\big) > R\big(\mathcal{F}_c\setminus i\big) + R\big(\mathcal{F}_{\bar{c}}\big)\!~.
\end{align}
\end{small}
\end{definition}
In other words, flow $i$ prefers to be a member of coalition $\mathcal{F}_c$ than $\mathcal{F}_{\bar{c}}$ under the condition of increasing the sum system throughput.

When flow $i\in \mathcal{N}$ performs a switching from $\mathcal{F}_c$ to $\mathcal{F}_{\bar{c}}$, the original partition $\mathcal{F}=\big\{\mathcal{F}_c,\mathcal{F}_{\bar{c}}\big\}$ is changed to the new partition $\mathcal{F}'=\big\{\mathcal{F}_c\setminus \{i\},\mathcal{F}_{\bar{c}}\bigcup \{i\}\big\}$. The switching can occur if and only if the preference order as defined in (\ref{F_c}) is satisfied.

In our proposed algorithm based on the coalition game, we first initialize the system by a random coalition partition with two coalitions $\mathcal{F}_c$ and $\mathcal{F}_{\bar{c}}$. For any flow $i\in \mathcal{N}$, we denote its present coalition by $\mathcal{F}_c$. If $\mathcal{F}_{\bar{c}}\succ_{i}\mathcal{F}_c$, there will be a switching of flow $i$ from $\mathcal{F}_c$ to $\mathcal{F}_{\bar{c}}$, and the current coalition partition $\mathcal{F}=\big\{\mathcal{F}_c,\mathcal{F}_{\bar{c}}\big\}$ will be updated to the new partition $\mathcal{F}'=\big\{\mathcal{F}_c\setminus \{i\},\mathcal{F}_{\bar{c}}\bigcup \{i\}\big\}$. In this mechanism, every flow $i$ can leave its current coalition and join the other coalition, given that the new coalition is strictly preferred based on the definition in (\ref{F_c}) and the flow can make a greater contribution to the system performance in terms of throughput in the new coalition. It's worth noting that our proposed algorithm based on coalition formation game needs to find a coalition structure that maximizes the total utility rather than the individual payoffs of the players.


The pseudo code of the proposed coalition game based algorithm is given in Algorithm~\ref{ALG1}, which contains a user association algorithm for maximizing system throughput, and a transmission scheduling algorithm for maximizing the number of users meeting QoS requirements. More specifically, Algorithm 1 first initializes the parameters, and collects the set of flows that need to be scheduled in the current time slot, as well as the QoS requirements of each flow. Secondly, the proposed algorithm divides all flows into two coalitions randomly. Then, take a flow as the target, calculate $R\big(\mathcal{F}_c\big)$, $R\big(\mathcal{F}_{\bar{c}}\big)$, $R\big(\mathcal{F}_c\setminus \{i\}\big)$ and $R\big(\mathcal{F}_{\bar{c}}\bigcup \{i\}\big)$. If the system sum rate when the current flow stays in the current coalition is greater than that when it is moved to another coalition, the current flow will stay in the current coalition, as shown in lines 6 and 7 of Algorithm 1. The user association phase of the proposed algorithm is not completed until all flows have made the above judgments in turn. In the transmission scheduling phase, the proposed algorithm arranges each flow from small to large according to its QoS requirements, finds the maximum $N_{suc}$ flows with the required time slots less than $M$, and then transmits them.

Firstly, according to the QoS requirements, the amount of time slots required by each flow are obtained. Then the user association either with BS or with MR is determined according to the coalition formation algorithm. After the convergence of the algorithm, the flows are sorted according to their required numbers of time slots to meet their QoS requests from small to large. By prioritizing the flows with smaller time slots requirement during transmission scheduling, the number of the flows that satisfy their QoS requests is maximized.

The condition to terminate the algorithm is that the partition converges to the Nash-stable partition. If $\forall i\in \mathcal{N}$, $i\in\mathcal{F}_c\subset\mathcal{F}$, $\mathcal{F}_c\succ_i \mathcal{F}_{\bar{c}}\bigcup \{i\}$, where $\mathcal{F}_{\bar{c}}\bigcup \mathcal{F}_c=\mathcal{F}$ and $\mathcal{F}_{\bar{c}}\bigcap \mathcal{F}_{c}=\emptyset$, then the partition $\mathcal{F}= \{\mathcal{F}_c, \mathcal{F}_{\bar{c}}\}$ is Nash-stable. Starting from any initial coalition structure $\mathcal{F}_{ini}$, the proposed coalition formation algorithm will always converge to a final partition $\mathcal{F}_{fin}$ that is Nash-stable, after a sequence of switching operations. We will give related theoretical proofs in Section~\ref{S5.4}.

We then analyze the complexity of Algorithm 1. In a specific time slot, the established train-ground communication system can be regarded as temporarily static, i.e., the coalition of each flow is determined (not necessarily the optimal situation). The computational complexity of the sum rate of the coalition formed by the flows associated with BS is ${\rm O}({n_B})$, and that of the coalition formed by the flows associated with MR is ${\rm O}({n_M})$. At this time, considering the decision operation shown in the line 6 of Algorithm 1, the complexity corresponding to the user association phase is ${\rm O}(N \cdot ({n_B} \cdot {n_M} + 1))$. Then, the computational complexity of directly sorting the QoS requirements corresponding to   data flows at one time is ${\rm O}({n_{ts}})$. Considering that the worst case of determining $N_{suc}$ requires $N$ times of judgment, the complexity in the data transmission phase is ${\rm O}({n_{ts}} + N)$. To sum up, the computational complexity of Algorithm 1 is ${\rm O}(N({n_B} + {n_M} + 2) + {n_{ts}})$.

\subsection{Theoretical Analysis}\label{S5.4}
\emph{1) Convergence}: We first examine the convergence of the proposed algorithm theoretically.

\emph{Theorem 1}: With an arbitrary initial partition $\mathcal{F}_{ini}$, after the limited switching operations, a final partition $\mathcal{F}_{fin}$ obtained by the proposed algorithm can always be found, and the coalitions within $\mathcal{F}_{fin}$ maintain a disjoint relationship.

\emph{Proof}: Users are divided into $U'$ disjoint subsets in the coalition structure (CS), and each subset represents a coalition. We exhaustively examine all the CSs of $U'$ users and obtain the set $S$ of all CSs, given by $S = \{ {S_1},{S_2},...,{S_{\left| S \right|}}\} $. Based on the Bell number ${B_k}$, the cardinality of $S$ for the $U'$ users can be expressed as
\begin{equation}
B(U') = \left\{ {\begin{array}{*{20}{c}}
{1~,~~~~~~~~~~~~~~~~~~~~~~~~~~~U'{\rm{ =  0,}}}\\
{\sum\limits_{k = 0}^{U' - 1} {\left( {\begin{array}{*{20}{c}}
{U' - 1}\\
k
\end{array}} \right)} {B_k}{\rm{ ,       ~~~~~}}U' \ge 1.{\rm{   }}}
\end{array}} \right.
\label{proof1}
\end{equation}

In Algorithm 1, the number of switch operations in each iteration is up to $N$, and each user either joins a new coalition or still stays in the original coalition. Thus, the whole number of partitions is $B(N)$, and since we preset the number of the BS and MRs, the number of partitions will be decreased based on the Bell number. As the number of partitions is bounded, we can conclude that after a certain number of switch operations, a final partition $\mathcal{F}_{fin}$ will be obtained.

\emph{2) Stability}: We next examine the Nash-stable structure of the proposed algorithm theoretically.

\begin{definition}[{Nash-stable Structure}]\label{Dnash}
If for all $i \in N$, $i \subseteq {{\cal F}_c} \subset {\cal F}$, ${{\cal F}_c}{ \succ _i}~{{\cal F}_{\bar c}}\bigcup {\{ i\} } $, for all ${{\cal F}_{\bar c}} \subset {\cal F}$, ${{\cal F}_c} \ne {{\cal F}_{\bar c}}$, a coalitional patition $C = \{ {C_0},{C_1},...,{C_{n - 1}},{C_n}\} $ is Nash-stable.
\end{definition}

\emph{Theorem 2}: The final partition $\mathcal{F}_{fin}$ obtained by the Algorithm 1 is Nash-stable.

\emph{Proof}: When the switching operation no longer occurs, the proposed algorithm based on the coalition game will terminate. At the same time, the optimal value of system throughput is obtained, and the final partition is Nash-stable. We have ${{\cal F}^ * }=\arg {\max _{\cal F}}R({\cal F})$, for all ${{\cal F}_c} \subset {\cal F}$, and ${{\cal F}^* } = \{ {{\cal F}_0}^* ,{{\cal F}_1}^* ,...,{{\cal F}_{n - 1}}^* ,{{\cal F}_n}^* \} $ is the Nash-stable coalition partition.

If the partition still has tendency of switching, the final partition $\mathcal{F}_{fin}$ has not been found yet, which is contradictory to $\mathcal{F}_{fin}$ is the final partition. Consequently, the final partition $\mathcal{F}_{fin}$ of the proposed algorithm has reached Nash-stable.

\begin{algorithm}[tp]
\caption{Coalition Game based Algorithm for User Association and Transmission
Scheduling}
\label{ALG1}
\begin{algorithmic}[1]
\REQUIRE The flow set $\mathcal{N}=\{i: 1\le i\le N\}$ and the corresponding QoS set $\mathcal{Q}=\{Q_i:1\le i\le N\}$, $M$.
\STATE Perform a random partition $\mathcal{F}_{ini}$ of $\mathcal{N}$;
\STATE Set the current partition $\mathcal{F}_{cur}=\mathcal{F}_{ini}$;
\REPEAT
\STATE Choose a flow $i\! \in\! \mathcal{N}$ in a pre-determined order, denote its coalition as $\mathcal{F}_c$ and the other coalition as $\mathcal{F}_{\bar{c}}$;
\STATE Get $R\big(\mathcal{F}_c\big)$, $R\big(\mathcal{F}_{\bar{c}}\big)$, $R\big(\mathcal{F}_c\setminus \{i\}\big)$ and $R\big(\mathcal{F}_{\bar{c}}\bigcup \{i\}\big)$;
\IF{$R\big(\mathcal{F}_c\big) + R\big(\mathcal{F}_{\bar{c}}\big) < R\big(\mathcal{F}_c\setminus i\big) + R\big(\mathcal{F}_{\bar{c}}\bigcup i\big)$}
\STATE Flow $i$ leaves $\mathcal{F}_c$ and join in $\mathcal{F}_{\bar{c}}$;
\STATE Update $\mathcal{F}_{\rm cur}\! =\!\big\{\mathcal{F}_c\setminus \{i\},\mathcal{F}_{c'}\bigcup \{i\}\big\}$;
\ENDIF
\UNTIL{$\mathcal{F}_{\rm cur}$ Converges to the Nash-stable partition $\mathcal{F}_{fin}$};
\STATE Get the number of time slots required for each flow according to its throughput requirement;
\STATE Sort the flows according to their required numbers of time slots from small to large;
\STATE Terminate when the first $N_{\rm suc}$ sorted flows have the number of time slots $\le M$ but counterpart of the first $N_{\rm suc}+1$ sorted flows $> M$;
\STATE \textbf{Output}: The first $N_{\rm suc}$ sorted flows.
\end{algorithmic}
\end{algorithm}

\section{Performance Evaluation}\label{S6}

In this section, we evaluate the performance of our proposed coalition game based algorithm under various system parameters. In the first part, we consider bandwidth resource allocation and ascertain the best ratio. In the second part, we compare our coalition game based scheme with other three schemes in terms of sum rate. After user association is over, we compare the proposed transmission scheduling scheme with another scheme in terms of system throughput and the number of users whose needs are met. Besides, we give the necessary analysis for the obtained simulation results.
\subsection{Simulation Setup}
Based on the system model described in Section~\ref{S3}, the entire train-ground communication system works in the 60GHz mm-wave frequency band, and we consider a single cell scenario where users are randomly distributed in a $300\times300$ square area. On the other hand, we use the widely used realistic directional antenna model in IEEE 802.15.3c~\cite{ant1}~\cite{ant2}, and the antenna gain in units of dB, denoted by ${G\left(\theta\right)}$ , can be expressed as
\begin{align}\label{An-G} 
G(\theta ) =& \left\{ \begin{array}{cl}
G_0 - 3.01 \left(\frac{2\theta}{\theta_{\rm -3dB}}\right)^2 , & ~0^{\circ} \le \theta \le \frac{\theta_{\rm ml}}{2} , \\
G_{\rm sl} ~, & \frac{\theta _{\rm ml}}{2} \le \theta  \le 180^{\circ} .
\end{array} \right.
\end{align}
where $\theta$ denotes an arbitrary angle within ${\left[0^{\circ},180^{\circ}\right]}$, $\theta _{-3dB}$ denotes the angle of the half-power beam width, and $\theta _{ml}$ is the main lobe width in units of degrees, and the relationship between $\theta _{ml}$ and $\theta _{-3dB}$ is ${\theta _{ml}=2.6 \cdot \theta _{-3dB}}$. $G_{\rm sl}$ is the side lobe gain given by $G_{\rm sl}\! =\! -0.4111$ $\times \ln\left(\theta_{\rm -3dB}\right) - 10.579$. $G_0$ is the maximum gain which can be obtained by
\begin{align}\label{An-G_0} 
 G_0 =& 10\log_{10}\left(\frac{1.6162}{\sin\big(\theta_{\rm -3dB}/2\big)}\right)^2 .
\end{align}

\begin{table}[t]
\centering
\caption{Simulation Parameters}
\scalebox{1.1}{
\begin{tabular}{l c c}
\hline
Parameter&Symbol&Value\\
\hline
System bandwidth & $W$ & 1GHz\\
Transmit power & $P_t^i$ & 30dBm\\
Background noise & $N_0$ & $-$174dBm/MHz\\
Bandwidth share of BS & $a$ & 0.4\\
Bandwidth share of MR & $b$ & 0.6\\
path loss exponent & $n$ & 3\\
SI cancellation level & $\beta$ & $10^{-15}$\\
\hline
\end{tabular}
}
\end{table}\par
Other simulations parameters are given in Table I. In order to show the advantage of the proposed coalition game based scheme, we compare our scheme, which is labeled as Coalition Game $\left(CG-FD\right)$, with other three schemes:
\subsubsection{CG-HD}
Coalition Game - Half Duplex~\cite{cghd}, where the MR operate in the half duplex mode. Thus, the bandwidth resource that the users can occupy is decreased in comparison with the case of FD under the same number of users.
\subsubsection{FBSC}
Full BS Communication~\cite{algref}, where all users are associated with the BS.
\subsubsection{FMRC}
Full MR Communication~\cite{algref}, where all users are associated with the MR.


\subsection{Rationality of the Model in the Reality}
Considering that HST mainly operates in rural areas and viaducts, and the buildings in the communication environments are generally low, so LOS transmission is dominant at this time. Mm-wave communications are directional, so we adopted the real directional antenna model in IEEE 802.15.3c~\cite{ant1}~\cite{ant2}. Secondly, we refer to the classical Shannon channel capacity for data rate. Thirdly, considering that the current algorithm can reduce the SI generated by FD to a low level, but there is still some residual SI in actual communication, a residual SI model such as $\beta \cdot P_t$ is built, and the results corresponding to different $\beta$ and $P_t$ on system capacity is analyzed in the following. Fourthly, considering the small coverage of the mm-wave BS, the radius of cell is set to 300m during simulation.

At present, train-ground wireless communication in HSR scenario mainly based on GSM-R, while mm-wave and FD are the key technologies of 5G. The details of their practical application in HSR scenarios are still unclear, so the construction of system models and the selection of simulation parameters mainly come from theoretical literatures and standards.


\subsection{Compared with the Optimal Solution}
\begin{figure}[t]
\begin{minipage}[t]{0.5\linewidth}
\centering
\includegraphics*[width=0.95\columnwidth,height=1.5in]{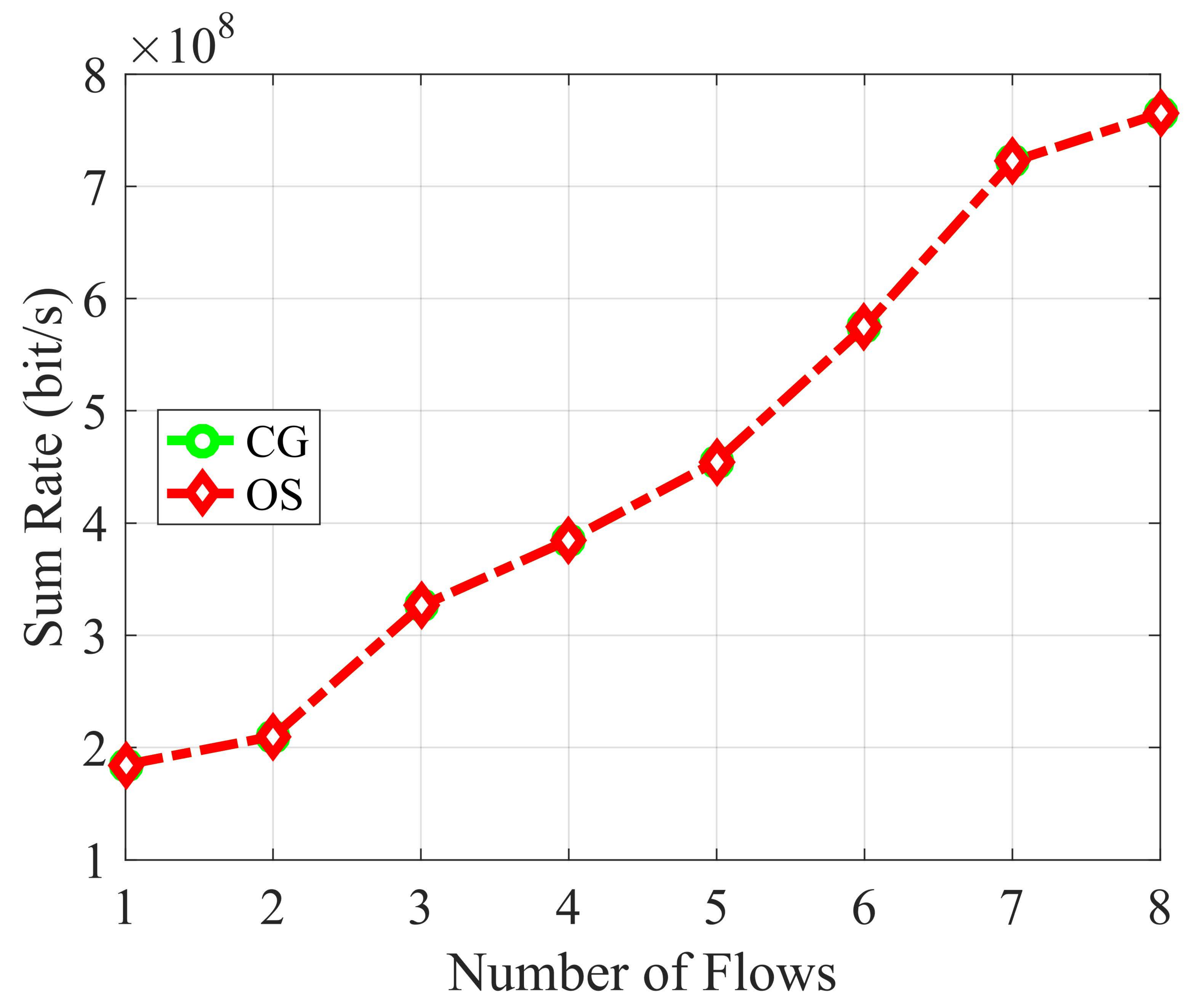}
\centerline{\small (a)}
\end{minipage}%
\begin{minipage}[t]{0.5\linewidth}
\centering
\includegraphics[width=1\columnwidth,height=1.5in]{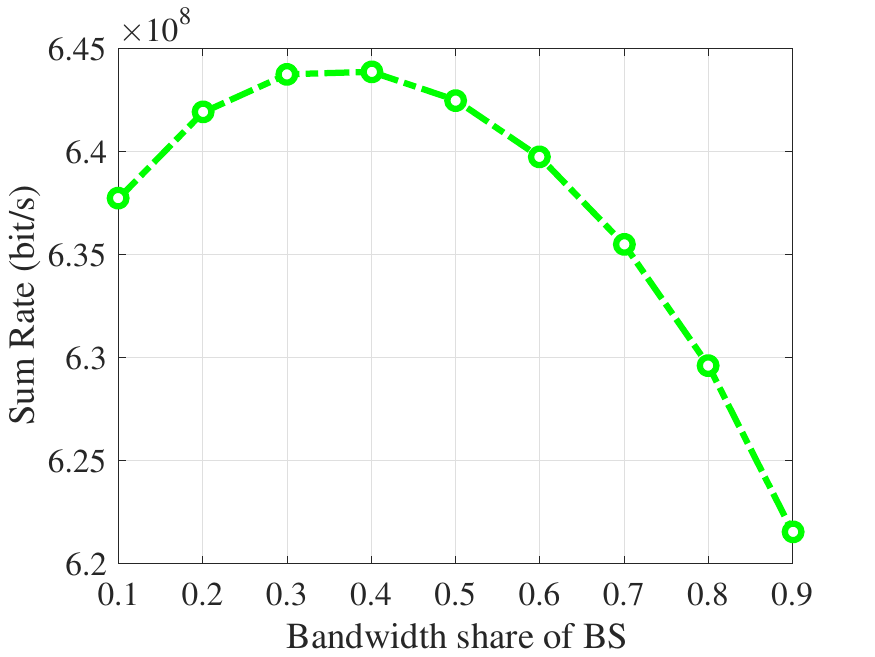}
\centerline{\small (b)}
\end{minipage}
\caption{Sum rate under different (a) number of flows and (b) bandwidth share of BS.}
\vspace*{-3mm}
\label{1and2}
\end{figure}

Firstly, we compare the performance of CG-FD with the optimal solution (OS) obtained by traditional exhaustive search method. The simulation results when the number of flows is 1 to 8 as shown in Fig.\ref{1and2} (a), and we can see that there is a good approximation between the sum rate of CG-FD shown by the solid line and that of OS shown by the dot line. This reminds us that under the same parameters, the user association algorithm based on coalition game proposed in this paper has the same optimal sum rate performance as the exhaustive search method, which is consistent with the theoretical analysis in Section~\ref{S4}.


\subsection{Bandwidth resource allocation}
In this section, we use simulation to determine the optimal way to allocate bandwidth. As shown in Fig.~\ref{1and2} (b), we obtain the sum rate with $a$, i.e., the ratio of bandwidth reserved by the BS to the total bandwidth resources, varying from 0.1 to 0.9.
From the simulation results, we can easily get that the optimal bandwidth allocation is 0.4 for the BS and 0.6 for the MR. At this time, the sum rate of the system reaches the maximum when other parameters remain unchanged. Therefore, in the following simulation and analysis section, we assume that the ratio of bandwidth to BS is 0.4 and the counterpart of MR is 0.6.

\subsection{System Sum Rate}


\begin{figure}[t]
\centering
\includegraphics*[width=0.95\columnwidth,height=2.65in]{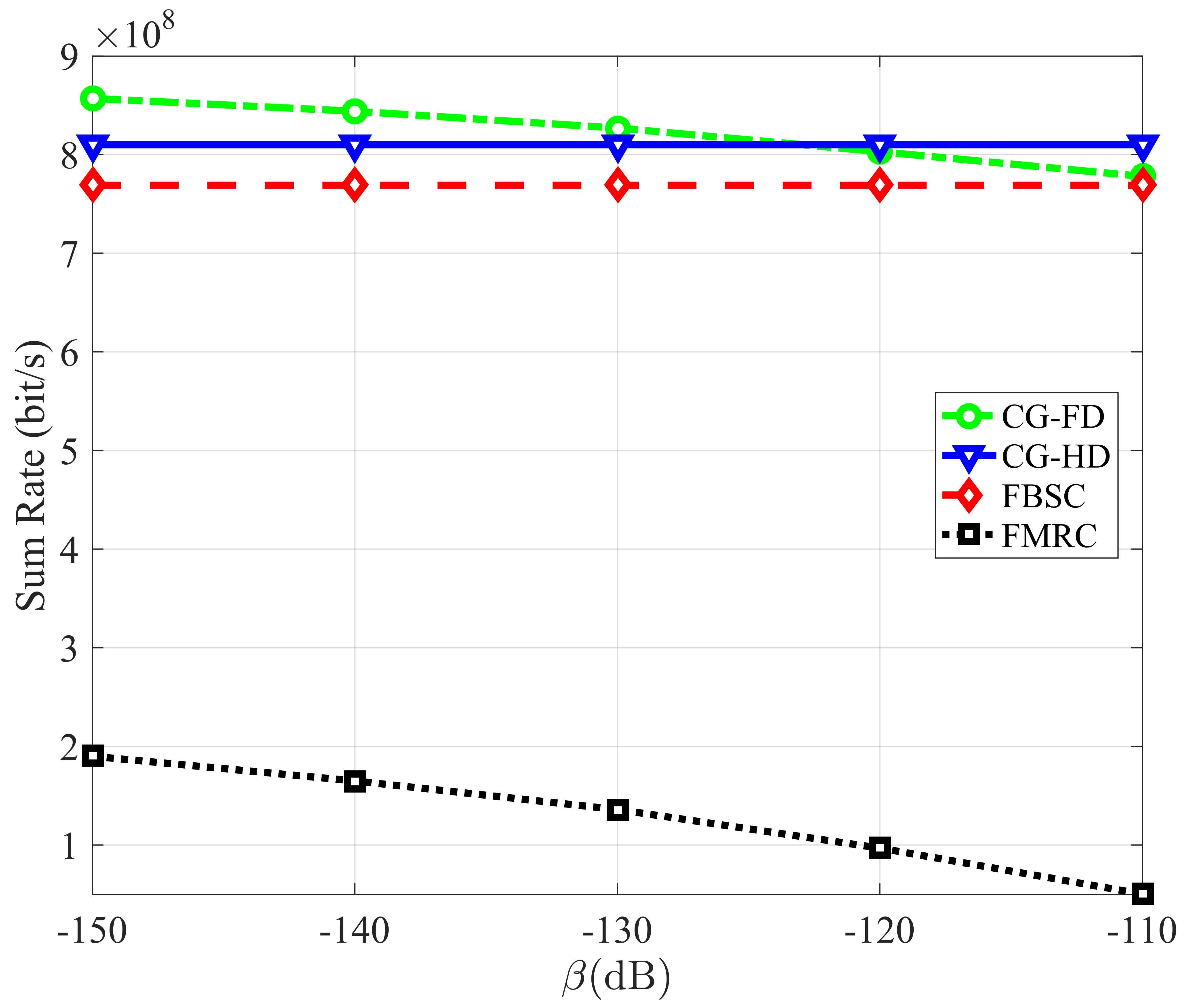}
\caption{Sum rate under different SI cancellation level.}\label{Beta_change}
\end{figure}

\begin{figure}[t]
\centering
\includegraphics*[width=0.95\columnwidth,height=2.65in]{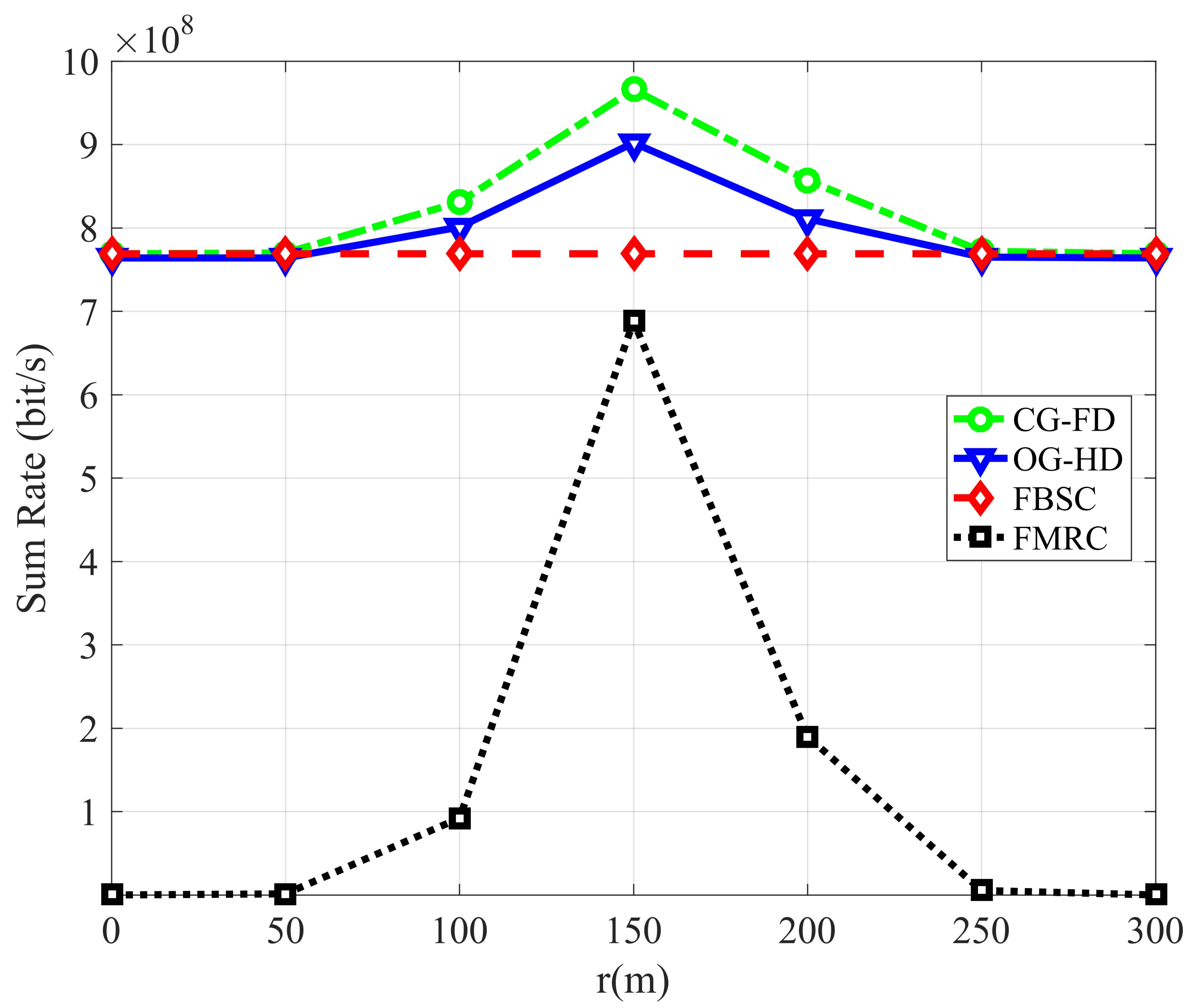}
\caption{Sum rate under different distance between BS and MR.}\label{Distance_change}
\end{figure}

\begin{figure}[t]
\centering
\includegraphics*[width=0.95\columnwidth,height=2.75in]{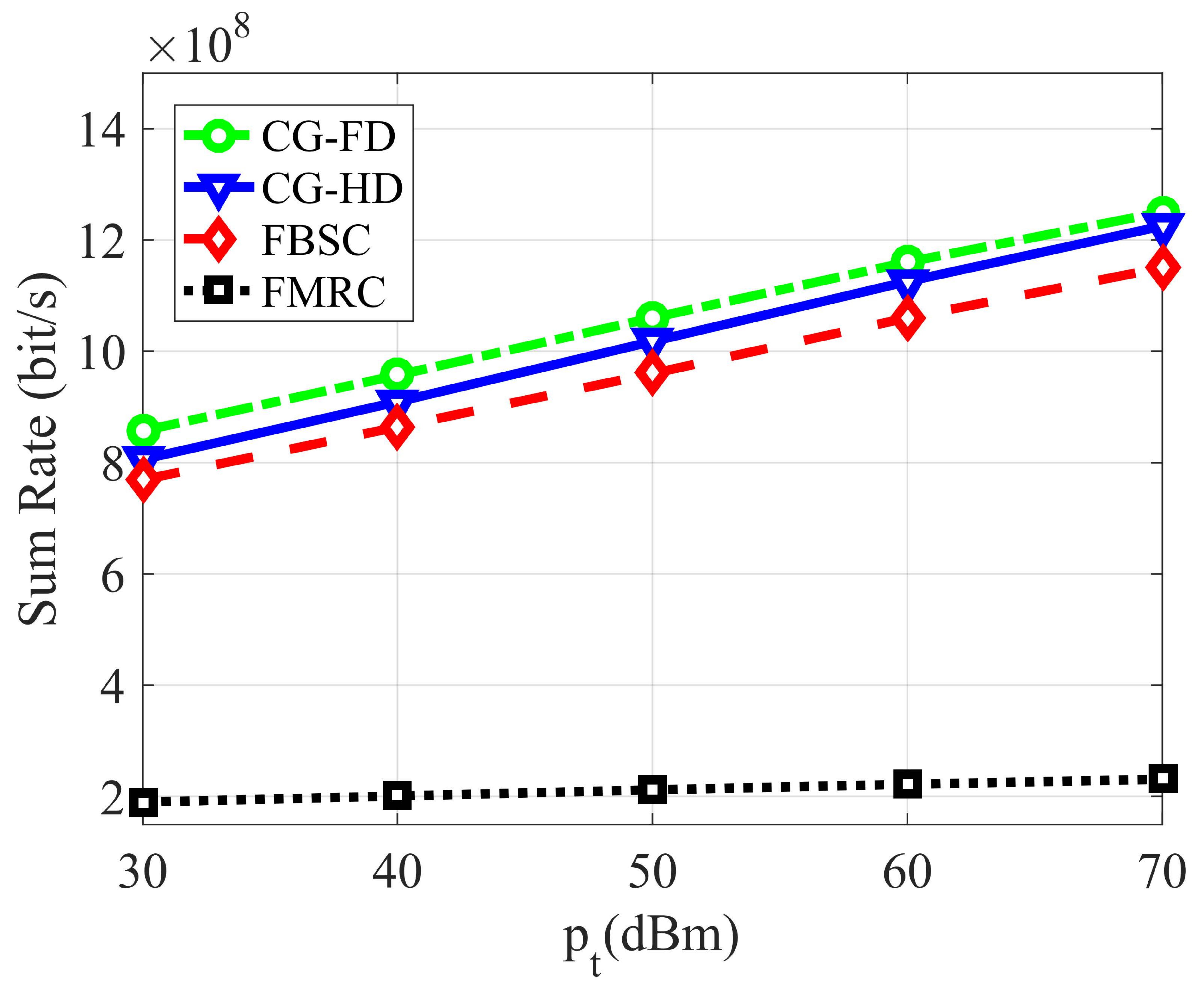}
\caption{Sum rate under different transmission power.}\label{Pt_change}
\end{figure}

When the number of flows waiting to be scheduled is 10, the sum rate obtained by the proposed coalition game based algorithm and other three comparison schemes when $\beta$ changes from $-$150 dB to $-$110 dB are shown in Fig.~\ref{Beta_change}. When $\beta=-150$ dB, the sum rate of CG-FD is larger than that of FBSC, FMRC and CG-HD about 11\%, 351\% and 6\%. As $\beta$ increases, the data rate of the flows transmitted from the MR decreases due to the increase of the SI, which can explain the trend of curves obtained by the CG-FD and the FMRC. Because there is no SI in the link without MR forwarding, the sum rate obtained by the FBSC scheme is stable. Unsurprisingly, as RSI gradually increases, more users intend to associate with the BS, so the curve corresponding to FMRC shows a trend of downward, which is obviously different from the other three curves. When $\beta$ is about -120dB, the sum rate obtained by CG-FD is lower than that obtained by CG-HD, which indicates when the SI cancellation level is above -120dB, the performance improvement brought by FD cannot offset the adverse effects of SI on the system.

As shown in Fig.~\ref{Distance_change}, we compare system sum rate of the four schemes with different distance between BS and MR. In the investigated mm-wave train-ground communication system, the longitudinal distance between the BS and MR is fixed, so we take the lateral displacement of the MR as the independent variable to simulate the operation of the train. Based on the simulation results, it can be seen that CG-FD has the best performance. When the lateral displacement of the MR is 150m, the sum rate of CG-FD is larger than that of FBSC, FMRC and CG-HD about 26\%, 40\% and 8\%, respectively. In general, when the distance between the BS and MR is too large, the sum rate of the MR is greatly limited, resulting in FMRC being the scheme with the lowest performance. On the whole, when the distance between BS and MR is about 150m, the sum rate have a maximum value. This is related to our planned radius of 300m. When other parameters are fixed and the distance between BS and MR is about 150m, the average distance between the user and the associated device is the shortest and the path loss is the smallest.

The variation of system sum rate under different transmission power is shown in Fig.~\ref{Pt_change}. When $P_t=30$dBm, the sum rate of CG-FD is larger than that of FBSC, FMRC and CG-HD about 11\%, 351\% and 6\%. As $P_t$ increases, the SI of MR also increases, so the curve of FMRC scheme has a flat trend. Since there is no SI in the flows associated with the BS, increasing $P_t$ is not bad for them and certainly has a greater effect on the growth of the FBSC. When the $P_t$ is low, the proposed CG-FD scheme is closer to the FMRC, and shows great advantages when $P_t$ is high.

\begin{figure}[t]
\centering
\includegraphics[width=0.95\columnwidth,height=2.685in]{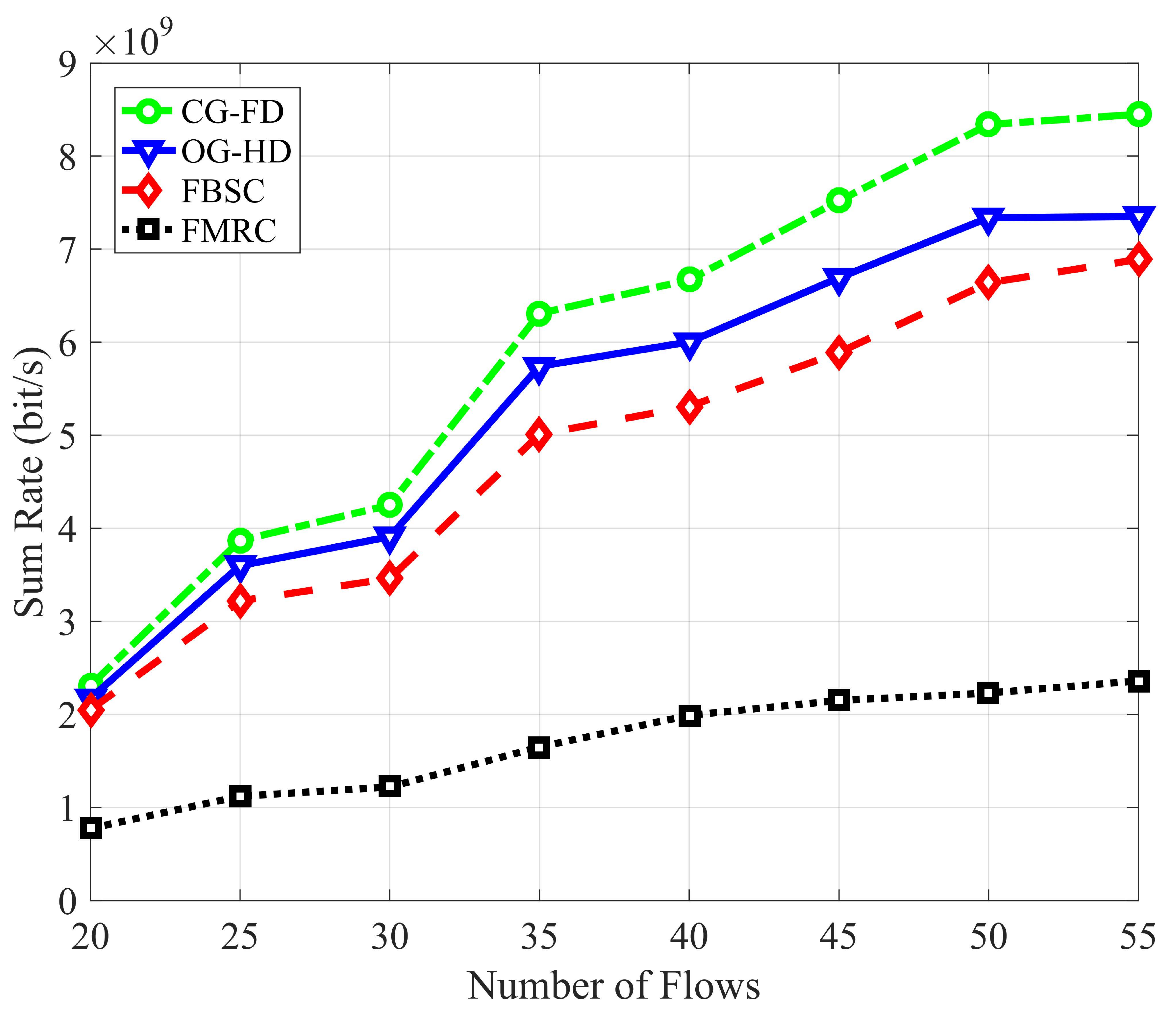}
\caption{Sum rate under different number of flows.}\label{Flow_change}
\end{figure}

As shown in Fig.~\ref{Flow_change}, we compare system sum rate of the four
schemes under different number of flows from 20 to 55. It can be seen that the performance of CG-FD proposed in this paper is best. When the number of the flows is 55, the sum rate of CG-FD is larger than that of FBSC, FMRC and CG-HD about 18\%, 254\% and 13\%. From overall view, the sum rate of these four schemes improve as the number of flows increases.

\begin{figure}[t]
\begin{minipage}[t]{0.5\linewidth}
\centering
\includegraphics[width=1.0\columnwidth,height=1.5in]{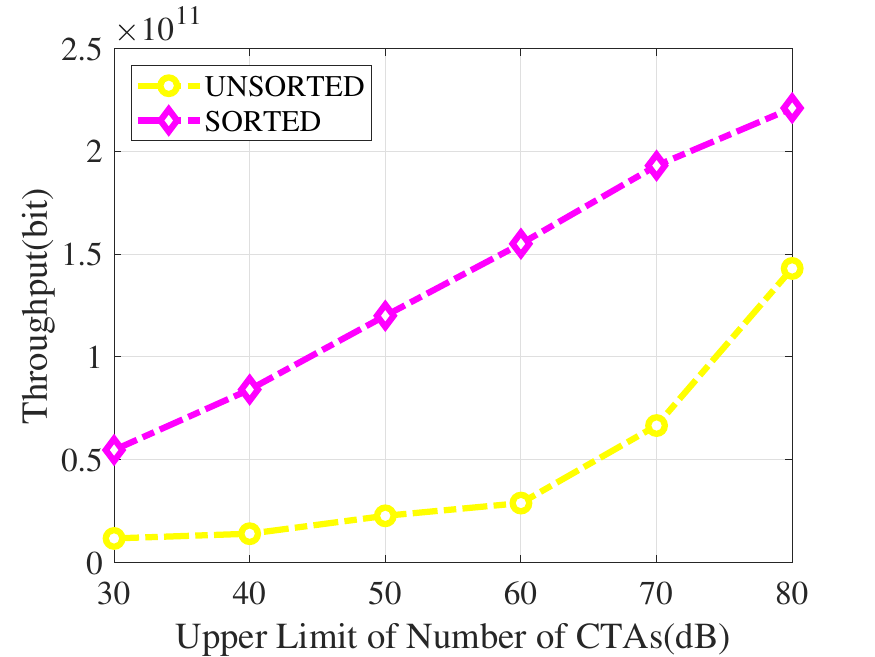}
\centerline{\small (a)}
\end{minipage}%
\begin{minipage}[t]{0.5\linewidth}
\centering
\includegraphics[width=1.0\columnwidth,height=1.5in]{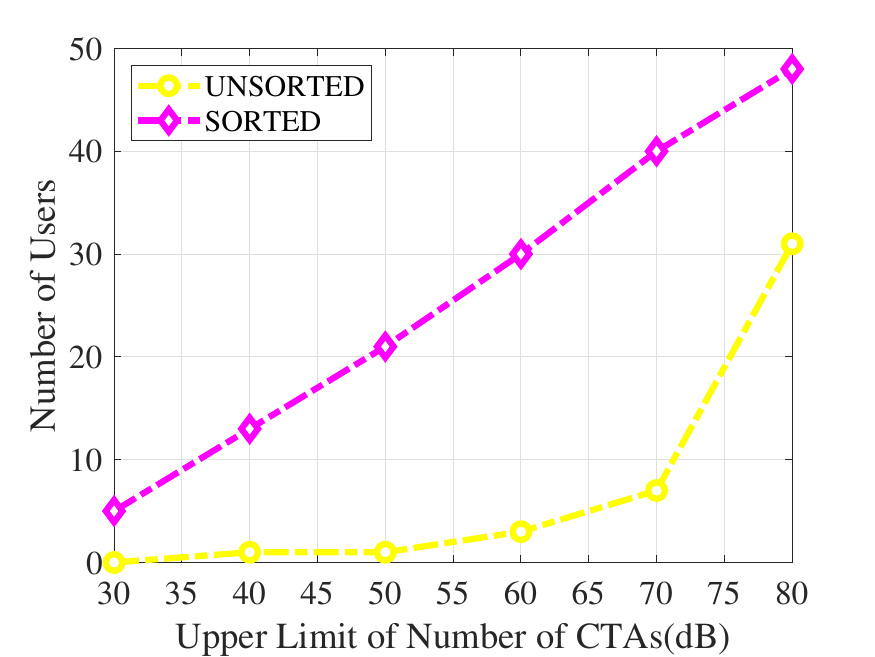}
\centerline{\small (b)}
\end{minipage}
\caption{The (a) system throughput and (b) number of users whose QoS requirements are met under different number of CTAs.}
\vspace*{-3mm}
\label{7and8}
\end{figure}

%

\begin{figure}[!t]
\centering
\includegraphics[width=0.8\columnwidth,height=2.25in]{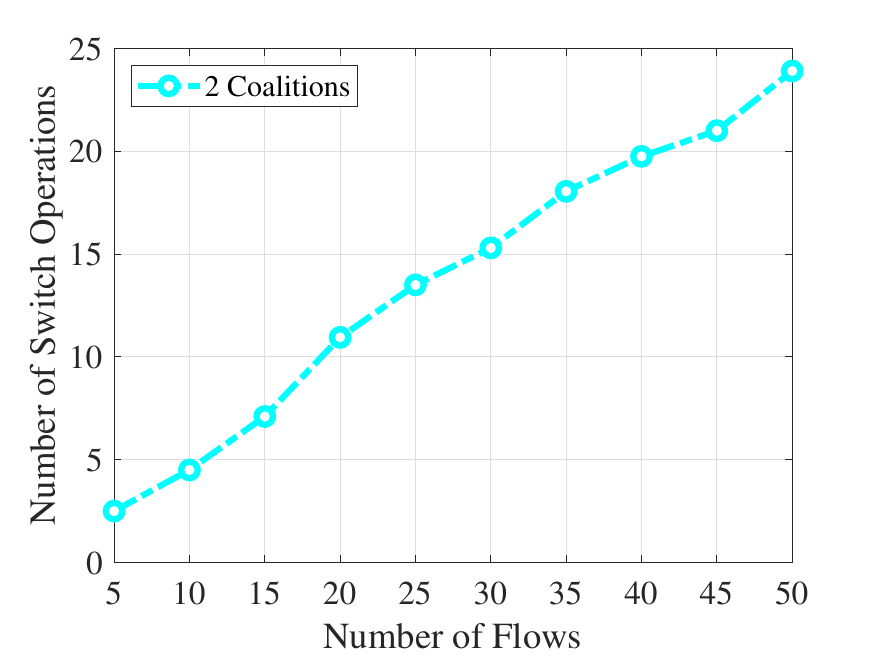}
\caption{Number of switch operations under different number of flows.}\label{Switch_operation}
\end{figure}
\subsection{System Performance}
In this simulation, we evaluate the system performance of our proposed transmission scheduling scheme. At first, we randomly set the throughput requirements of each flow and the upper limit of the required number of time slots. The simulation results of system performance under different number of CTAs are shown in Fig.~\ref{7and8}.

When the total number of flows is 50, the system throughput under different upper limit of number of time slots are shown in Fig.~\ref{7and8} (a). We can see that the proposed transmission scheduling scheme produces higher throughput and better performance. After sorting the number of slots required by all the flows, the proposed scheme were able to achieve much higher throughput than the one without sorting with limited slots, thus improving the performance of the system.

The variation of number of users whose throughput requirements are met under different restrictions of time slots are shown in Fig.~\ref{7and8} (b). We can see that the performance of the proposed transmission scheduling scheme is better. By sorting the number of slots required by all flows, we are able to meet the needs of more users with a limited number of time slots.
\subsection{Convergence Rate}
To examine the convergence performance of the proposed coalition game based algorithm, the number of switch operations under different number of flows are shown in Fig.~\ref{Switch_operation} where the proposed CG scheme converges to the final partition. As we can see, as the number of flows increases, so does the number of switching operations, and the average number of switching operations is 0 to 25. Given $N$ flows, the exhaustive operation requires $2^N$ iterations to find the optimal solution. Therefore, the proposed coalition game based algorithm allows users to establish the final Nash-stable partition with a very fast convergence rate, which greatly reduces the computational complexity.


\subsection{Challenges in Real-life Environments}
The above results show that the proposed algorithm can effectively improve the system throughput while maintaining a certain QoS awareness, but there are still some challenges in applying them in real-life environments.

Firstly, the established train-ground communication system works in the mm-wave band, which requires the antenna array to be arranged at MR and BS simultaneously. Considering the inherent high path loss of mm-wave communications, we also need to optimize the beamforming at the transmitter and receiver to maximize the directional gain of the antenna. At present, the method based on deep reinforcement learning has been proposed, which has great potential to solve this problem.

Secondly, considering that the operation of HST is based on a fixed timetable and its trajectory is predictable, the track-side train-ground wireless communication system also has a regular pattern, and artificial intelligence is good at capturing this hidden pattern, which has great potential to optimize the beamforming in the HSR scenarios~\cite{xutao}~\cite{AIenable}. In addition, as the HST mainly runs in rural areas and viaducts, the buildings in the communication environments are generally low. At this time, LOS transmission is dominant, but there is still the possibility that the service may be interrupted due to blocking. At present, scholars have proposed to deploy IRS on HST windows or track-side~\cite{gaoIRS}.

Thirdly, in the actual scenarios, BS should timely count the QoS requirements of users, and send the results of user association and transmission scheduling to users through the control channel. However, because the speed of HST is fast and not constant, there is usually additional queuing and transmission delay, so a more detailed model needs to be established to further evaluate the control channel bandwidth that needs to be reserved when the HST speed is different, and the implementation complexity of the algorithm can also be reduced by referring to historical user association and transmission scheduling data.

\section{Conclusions}\label{S7}

In this paper, we propose a coalition game based algorithm to solve the user association and transmission scheduling problems in the mm-wave train-ground communication system with MR operating in the FD mode. The coalition formation algorithm ensures the maximum sum rate of the system, and the sorting algorithm ensures the maximum number of users meeting QoS requirements. In addition, we theoretically proved that the proposed algorithm can converge after limited steps, and finally reach the Nash-stable structure. Extensive simulation results demonstrated that the proposed coalition game based algorithm can effectively improve the system throughput while maintaining a certain QoS awareness, when compared with three baseline schemes.


\bibliographystyle{IEEEtran}

\begin{thebibliography}{10}
\providecommand{\url}[1]{#1}
\csname url@samestyle\endcsname
\providecommand{\newblock}{\relax}
\providecommand{\bibinfo}[2]{#2}
\providecommand{\BIBentrySTDinterwordspacing}{\spaceskip=0pt\relax}
\providecommand{\BIBentryALTinterwordstretchfactor}{4}
\providecommand{\BIBentryALTinterwordspacing}{\spaceskip=\fontdimen2\font plus
\BIBentryALTinterwordstretchfactor\fontdimen3\font minus
  \fontdimen4\font\relax}
\providecommand{\BIBforeignlanguage}[2]{{%
\expandafter\ifx\csname l@#1\endcsname\relax
\typeout{** WARNING: IEEEtran.bst: No hyphenation pattern has been}%
\typeout{** loaded for the language `#1'. Using the pattern for}%
\typeout{** the default language instead.}%
\else
\language=\csname l@#1\endcsname
\fi
#2}}
\providecommand{\BIBdecl}{\relax}
\BIBdecl
\bibitem{HSR2018} 
\emph{High Speed Rail Development Worldwide}, EESI Fact Sheet, June 2018. [Online]. Available:
https://www.eesi.org/papers/view/fact-sheet-high-speed-rail-development-worldwide

\bibitem{Ai}
B. Ai, A. F. Molisch, M. Rupp and Z. Zhong, ``5G key technologies for smart railways," in Proceedings of the IEEE, vol. 108, no. 6, pp. 856-893, June 2020.

\bibitem{Doppler}
D. Fan, Z. Zhong, G. Wang, and F. Gao, ``Doppler shift estimation for high-speed railway wireless communication systems with large-scale linear antennas," in Proc. Int. Workshop High Mobility Wireless Commun., pp. 96-100, Oct. 2015.

\bibitem{HSRMR}
M. Gao et al., ``Dynamic mmWave beam tracking for high speed railway communications," in Proc. IEEE Wireless Commun. Netw. Conf. Workshops (WCNCW), pp. 278-283, Apr. 2018.

\bibitem{xiangfei21} 
X. Zhang et al., ``Resource Allocation for Millimeter-Wave Train-Ground Communications in High-Speed Railway Scenarios," in IEEE Transactions on Vehicular Technology, vol. 70, no. 5, pp. 4823-4838, May 2021.

\bibitem{ROF}
B. Lannoo, D. Colle, M. Pickavet, and P. Demeester, ``Radio-over-fiber-based solution to provide broadband Internet access to train passengers,"
{\em IEEE Communications Magazine}, vol. 45, no. 2, pp. 56-62, Feb. 2007.

\bibitem{J1}
Y.~Niu, C.~Gao, Y.~Li, D.~Jin, L.~Su, and D.~Wu, ``Boosting spatial reuse via
  multiple paths multi-hop scheduling for directional mm{W}ave {WPANs},''
  \emph{IEEE Transactions on Vehicular Technology}, vol.~65, no.~8, pp.
  6614--6627, Aug. 2016.

\bibitem{channel}
K. Guan et al., ``On millimeter wave and thz mobile radio channel for smart rail
  mobility,'' \emph{IEEE Transactions on Vehicular Technology}, vol.~66, no.~7,
  pp. 5658--5674, Jul. 2017.

\bibitem{J8}
H.~{Song}, X.~{Fang}, and Y.~{Fang}, ``Millimeter-wave network architectures
  for future high-speed railway communications: Challenges and solutions,''
  \emph{IEEE Wireless Communications}, vol.~23, no.~6, pp. 114--122, Dec. 2016.

\bibitem{24ghz}
S. Singh, R. Mudumbai, and U. Madhow, ``Interference analysis for highly directional 60-GHz mesh networks: the case for rethinking medium access control," IEEE/ACM Transactions on Networking (TON), vol. 19, no. 5, pp. 1513-1527, Oct. 2011.

\bibitem{fd1}
W. G. Ding, Y. Niu, H. Wu, Y. Li, and Z. D. Zhong, ``QoS-Aware Full-Duplex Concurrent Scheduling for Millimeter Wave Wireless Backhaul Networks," IEEE Access, vol. 6, pp. 25313-25322, 2018.

\bibitem{fd2}
C. Skouroumounis, C. Psomas, and I. Krikidis, ``Heterogeneous FD mm-Wave Cellular Networks With Cell Center/Edge Users," IEEE Transactions on Communications, vol. 67, no. 1, pp. 791-806, Jan. 2019.

\bibitem{FDintr}
Z. Zhang, X. Chai, K. Long, A. V. Vasilakos and L. Hanzo, ``Full duplex techniques for 5G networks: self-interference cancellation, protocol design, and relay selection," {\em IEEE Communications Magazine}, vol. 53, no. 5, pp. 128-137, May 2015.

\bibitem{LuExploring2018} 
Y.~Z.~Lu, J.~H.~Qi, F.~Q.~Hu, and H.~Peng, ``Exploring the application of millimeter wave communication technology in high-speed rail passenger service,'' \emph{Railway Signaling Commun.}, vol.~54, no.~5, pp.~50--53, 2018.

\bibitem{DoanDesign2004} 
C.~H.~Doan, \emph{et al.}, ``Design considerations for 60 GHz CMOS radios,'' \emph{IEEE Commun. Mag.}, vol.~42, no.~12, pp.~132--140, Dec. 2004.

\bibitem{kt15}
S. Singh, R. Mudumbai and U. Madhow, ``Interference Analysis for Highly Directional 60-GHz Mesh Networks: The Case for Rethinking Medium Access Control," in IEEE/ACM Transactions on Networking, vol. 19, no. 5, pp. 1513-1527, Oct. 2011.

\bibitem{kt14}
S. Singh, F. Ziliotto, U. Madhow, E. Belding and M. Rodwell, ``Blockage and directivity in 60 GHz wireless personal area networks: from cross-layer model to multihop MAC design," in IEEE Journal on Selected Areas in Communications, vol. 27, no. 8, pp. 1400-1413, October 2009.

\bibitem{kt18}
ETSI TS 136 216-2018,Universal Mobile Telecommunications System (UMTS); LTE; Evolved Universal Terrestrial Radio Access (E-UTRA); Physical layer for relaying operation (V15.0.0; 3GPP TS 36.216 version 15.0.0 Release 15).

\bibitem{kt19}
A. Khan and A. Jamalipour, ``Moving Relays in Heterogeneous Cellular Networks?A Coverage Performance Analysis," in IEEE Transactions on Vehicular Technology, vol. 65, no. 8, pp. 6128-6135, Aug. 2016.

\bibitem{kt20}
A. J. Theiss, C. J. Meadows and R. B. True, ``Experimental Investigation of a Novel Circuit for MM-Wave TWTs," 2006 IEEE International Vacuum Electronics Conference held Jointly with 2006 IEEE International Vacuum Electron Sources, pp. 183-184, 2006.

\bibitem{kt23}
D. He et al., ``Influence of Typical Railway Objects in a mmWave Propagation Channel," in IEEE Transactions on Vehicular Technology, vol. 67, no. 4, pp. 2880-2892, April 2018.

\bibitem{refr21}
H. Yuan and M. Zhou, "Profit-Maximized Collaborative Computation Offloading and Resource Allocation in Distributed Cloud and Edge Computing Systems," in IEEE Transactions on Automation Science and Engineering, vol. 18, no. 3, pp. 1277-1287, July 2021.

\bibitem{FDchen}
X. Chen et al., ``Multi-Antenna Covert Communication via Full-Duplex Jamming Against a Warden With Uncertain Locations," in IEEE Transactions on Wireless Communications, vol. 20, no. 8, pp. 5467-5480, Aug. 2021.

\bibitem{FDcao}
Y. Cao et al., ``Secrecy Analysis for Cooperative NOMA Networks With Multi-Antenna Full-Duplex Relay," in IEEE Transactions on Communications, vol. 67, no. 8, pp. 5574-5587, Aug. 2019.

\bibitem{FDli}
L. Weizheng and T. Xiumei, ``Throughput analysis of full-duplex network coding in two-way relay channel," 2017 IEEE 17th International Conference on Communication Technology (ICCT), 2017, pp. 85-90.

\bibitem{FDcheng}
W. Cheng, X. Zhang and H. Zhang, ``RTS/FCTS mechanism based full-duplex MAC protocol for wireless networks," 2013 IEEE Global Communications Conference (GLOBECOM), 2013, pp. 5017-5022.

\bibitem{FDkai}
C. Kai, X. Zhang, X. Hu and W. Huang, ``Optimal Scheduling and Power Control for In-Band Full-Duplex Communication in WLANs," GLOBECOM 2020 - 2020 IEEE Global Communications Conference, 2020, pp. 1-6.

\bibitem{WenTime2016} 
D.~Wen and G.~Yu, ``Time-division cellular networks with full-duplex base stations,'' \emph{IEEE Commun. Lett.}, vol.~20, no.~2, pp. 392--395, Feb. 2016.

\bibitem{DingQoS2018} 
W.~Ding, \emph{et al.}, ``QoS-aware full-duplex concurrent scheduling for millimeter wave wireless backhaul networks,'' \emph{IEEE Access}, vol.~6, pp.~25313--25322, Apr. 2018.

\bibitem{NiuA2015} 
Y.~Niu, \emph{et al.}, ``A survey of millimeter wave communications (mmWave) for 5G: Opportunities and challenges,'' \emph{Wireless Networks}, vol.~6, no.~8, pp.~2657--2676, 2015.

\bibitem{FDnidal}
N. Zarifeh, M. Alissa, M. Khaliel and T. Kaiser, ``Self-interference mitigation in full-duplex base-station using dual polarized reflect-array," 2018 11th German Microwave Conference (GeMiC), 2018, pp. 180-183.

\bibitem{refr22}
Y. Du, L. Wang, L. Xing, J. Yan and M. Cai, ``Data-Driven Heuristic Assisted Memetic Algorithm for Efficient Inter-Satellite Link Scheduling in the BeiDou Navigation Satellite System," in IEEE/CAA Journal of Automatica Sinica, vol. 8, no. 11, pp. 1800-1816, Nov 2021.

\bibitem{refr23}
X. Jin, C. Xia, N. Guan and P. Zeng, ``Joint Algorithm of Message Fragmentation and No-Wait Scheduling for Time-Sensitive Networks," in IEEE/CAA Journal of Automatica Sinica, vol. 8, no. 2, pp. 478-490, Feb 2021.

\bibitem{refr24}
M. Taheri, N. Ansari, J. Feng, R. Rojas-Cessa and M. Zhou, "Provisioning Internet Access Using FSO in High-Speed Rail Networks," in IEEE Network, vol. 31, no. 4, pp. 96-101, July-August 2017.

\bibitem{comalg}
S. Rezvani, N. M. Yamchi, M. R. Javan and E. A. Jorswieck, ``Resource Allocation in Virtualized CoMP-NOMA HetNets: Multi-Connectivity for Joint Transmission," in IEEE Transactions on Communications, vol. 69, no. 6, pp. 4172-4185, June 2021.

\bibitem{LiConcurrent2017} 
Y.~Li, Z.~Zhang, W.~Wang, and H.~Wang, ``Concurrent transmission based Stackelberg game for D2D communications in mmWave networks,'' in \emph{Proc. ICC 2017} (Paris, France), May~21-25, 2017, pp.~1--6.

\bibitem{GAMEpaik}
S. H. Paik, S. Kim and H. B. Park, ``A resource allocation using game theory adopting AMC scheme in multi-cell OFDMA system," 2010 2nd International Conference on Future Computer and Communication, 2010, pp. V2-344-V2-347.

\bibitem{GAMEwang}
A. Wang, Y. Cai and Z. Hou, ``A novel resource allocation algorithm in uplink multi-cell OFDMA networks based on game theory," 2011 International Conference on Wireless Communications and Signal Processing (WCSP), 2011, pp. 1-4.

\bibitem{GAMErathi}
R. Rathi and N. Gupta, ``Device to Device Communication using Stackelberg Game Theory approach," 2020 Research, Innovation, Knowledge Management and Technology Application for Business Sustainability (INBUSH), 2020, pp. 100-103.

\bibitem{SonOn2012} 
I.~K.~Son, S.~Mao, M.~X.~Gong, and Y.~Li, ``On frame-based scheduling for directional mmWave WPANs,'' in \emph{Proc. INFOCOM 2012} (Orlando, FL, USA), Mar.~25-30, 2012, pp.~2149--2157.

\bibitem{CaiRex2010} 
L.~X.~Cai, L.~Cai, X.~Shen, and J.~W.~Mark, ``Rex: A randomized EXclusive region based scheduling scheme for mmWave WPANs with directional antenna,'' \emph{IEEE Trans. Wireless Commun.}, vol.~9, no.~1, pp.~113--121, Jan. 2010.

\bibitem{QiaoSTDMA2012} 
J.~Qiao, L.~X.~Cai, X.~Shen, and J.~W.~Mark, ``STDMA-based scheduling algorithm for concurrent transmissions in directional millimeter wave networks,'' in \emph{Proc. ICC 2012} (Ottawa, ON, Canada), Jun.~10-15, 2012, pp.~5221--5225.

\bibitem{blockconstant}
G. Yang, J. Du and M. Xiao, ``Maximum throughput path selection with random blockage for indoor 60 GHz relay networks," {\em IEEE Transactions on Communications}, vol. 63, no. 10, pp. 3511-3524, Oct. 2015.

\bibitem{01pack}
D. Pisinger, ``Where are the hard knapsack problems?" {\em Computers \& Operations Research}, vol. 32, no. 9, pp. 2271-2284, 2005.

\bibitem{Chen2012} 
Q.~Chen, X.~Peng, J.~Yang, and F.~Chin, ``Spatial reuse strategy in mmWave WPANs with directional antennas,'' in \emph{Proc. GLOBECOM 2012} (Anaheim, CA, USA), Dec.~3-7, 2012, pp.~5392--5397.

\bibitem{SaadCoalitional2009} 
W.~Saad, \emph{et al.}, ``Coalitional game theory for communication networks,'' \emph{IEEE Signal Processing Mag.}, vol.~26, no.~5, pp.~77--97, Sep. 2009.

\bibitem{ant1}
I. Toyoda, T. Seki, K. Iigusa,et al., ``Reference antenna model with side
lobe for TG3c evaluation," IEEE 802.15-06-0474-00-003c, Nov. 2006.

\bibitem{ant2}
Y. Niu et al., ``Energy-Efficient Scheduling for mmWave Backhauling of Small Cells in Heterogeneous Cellular Networks," in IEEE Transactions on Vehicular Technology, vol. 66, no. 3, pp. 2674-2687, March 2017.

\bibitem{cghd}
Y. Chen, Y. Niu, B. Ai, Z. Zhong, D. Wu and K. Li, ``Using Coalition Games for QoS Aware Scheduling in mmWave WPANs," 2018 IEEE 87th Vehicular Technology Conference (VTC Spring), 2018.

\bibitem{algref}
Study on new radio access technology:Radio Frequency (RF) and coexistence aspects, 3GPP TR 38.803, 2017.

\bibitem{xutao}
X. Zhou et al., ``Deep Reinforcement Learning Coordinated Receiver Beamforming for Millimeter-Wave Train-Ground Communications," in IEEE Transactions on Vehicular Technology, vol. 71, no. 5, pp. 5156-5171, May 2022.

\bibitem{AIenable}
L. Yan, X. Fang, X. Wang and B. Ai, ``AI-Enabled Sub-6-GHz and mm-Wave Hybrid Communications: Considerations for Use With Future HSR Wireless Systems," in IEEE Vehicular Technology Magazine, vol. 15, no. 3, pp. 59-67, Sept. 2020.

\bibitem{gaoIRS}
M. Gao, B. Ai, Y. Niu, Z. Han and Z. Zhong, ``IRS-Assisted High-Speed Train Communications: Outage Probability Minimization with Statistical CSI," ICC 2021 - IEEE International Conference on Communications, 2021.

%
%
%
%

%
%
%

%
%
\bibitem{CaoEnhancing2019} 
Y.~Cao, \emph{et al.}, ``Enhancing video QoE over high-speed train using segment-based prefetching and caching,'' \emph{IEEE MultiMedia}, vol.~26, no.~4, pp.~55--66, Oct.-Dec. 2019.

\bibitem{LiuExploiting2018} 
Y.~Liu, \emph{et al.}, ``Exploiting multi-hop relay to achieve mobility-aware transmission scheduling in mmWave systems,'' in \emph{Proc. ICC 2018} (Kansas City, MO, USA), May~20-24, 2018, pp.~1--6.

\bibitem{WanTime2017} 
X.~Wan, X.~Guan, J.~Wang, and M.~Zhao, ``Time slotted scheduling for outdoor mmWave mesh networks with performance guarantee,'' in \emph{Proc. ICNC 2017} (Silicon Valley, CA, USA), Jan.~26-29, 2017, pp.~224--229.

\bibitem{NiuEnergy2017} 
Y.~Niu, \emph{et al.}, ``Energy-efficient scheduling for mmWave backhauling of small cells in heterogeneous cellular networks,'' \emph{IEEE Trans. Vehicular Technology}, vol.~66, no.~3, pp.~2674--2687, Mar. 2017.

\bibitem{NiuExploiting2015} 
Y.~Niu, \emph{et al.}, ``Exploiting device-to-device communications in joint scheduling of access and backhaul for mmWave small cells,'' \emph{IEEE J. Sel. Areas Commun.}, vol.~33, no.~10, pp.~2052--2069, Oct. 2015.

\bibitem{ZhuRegret2017} 
Y.~Zhu J.~Li, Q.~Huang, and D.~Wu, ``Regret benefit ratio link scheduler for wireless backhaul with directional antennas,'' \emph{IEEE Trans. Vehicular Technology}, vol.~66, no.~11, pp.~10220--10232, Nov. 2017.

\bibitem{LiPractical2017} 
J.~Li, Y.~Zhu, and D.~O.~Wu, ``Practical distributed scheduling for QoS-aware small cell mmWave mesh backhaul network,'' \emph{Ad Hoc Networks}, vol.~55, pp.~62--71, Feb. 2017.

\end{thebibliography}

\bibliographystyle{IEEEtran}

\end{document}